\documentclass[a4paper,UKenglish,cleveref, autoref, thm-restate]{lipics-v2021}
\usepackage{siunitx}
\usepackage{tikz}
\usepackage[disable]{todonotes}
\usepackage[section]{algorithm}
\usepackage{mathtools}
\usepackage{algpseudocode}    
\usepackage{pseudo}
\usetikzlibrary{calc,patterns,angles,quotes}
\usetikzlibrary{decorations}
\usepackage{pgfplots}
\pgfplotsset{compat=1.18}

\pseudoset{label=\protect\arabic*.,  line-height=1.2, ref}

\DeclarePairedDelimiter{\ceil}{\lceil}{\rceil}

\pdfoutput=1 
\hideLIPIcs  

\newcommand{\est}{\textsc{EST}}
\newcommand{\esl}{\textsc{ESL}}
\newcommand{\esfl}{\textsc{ESfL}}

\title{NP-hardness and a PTAS for the Euclidean Steiner Line Problem} 

\author{Simon Bartlmae}{University of Bonn, Germany}{bartlmae@uni-bonn.de}{https://orcid.org/0009-0009-6953-8347}{}
\author{Paul J. Jünger}{University of Bonn, Germany}{s94pjuen@uni-bonn.de}{https://orcid.org/0009-0008-4165-4453}{}
\author{Elmar Langetepe}{University of Bonn, Germany}{Elmar.Langetepe@informatik.uni-bonn.de}{}{}

\authorrunning{S. Bartlmae, P. J. Jünger, and E. Langetepe} 

\Copyright{Simon Bartlmae, Paul J. Jünger, and Elmar Langetepe} 



\category{} 

\relatedversion{} 

\acknowledgements{We would like to thank Andreas Hene and Michael Kaibel for their valuable advice on improving the presentation of our results.}

\nolinenumbers 

\EventEditors{John Q. Open and Joan R. Access}
\EventNoEds{2}
\EventLongTitle{42nd Conference on Very Important Topics (CVIT 2016)}
\EventShortTitle{CVIT 2016}
\EventAcronym{CVIT}
\EventYear{2016}
\EventDate{December 24--27, 2016}
\EventLocation{Little Whinging, United Kingdom}
\EventLogo{}
\SeriesVolume{42}
\ArticleNo{23}

\begin{document}

\maketitle

\begin{abstract}
The Euclidean Steiner Tree Problem (\est) seeks a minimum-cost tree interconnecting a given set of terminal points in the Euclidean plane, allowing the use of additional intersection points. In this paper, we consider two variants that include an additional straight line $\gamma$ with zero cost, which must be incorporated into the tree. In the Euclidean Steiner fixed Line Problem (\esfl), this line is given as input and can be treated as a terminal. In contrast, the Euclidean Steiner Line Problem (\esl) requires determining the optimal location of $\gamma$.
Despite recent advances, including heuristics and a 1.214-approximation algorithm for both problems, a formal proof of NP-hardness has remained open.
In this work, we close this gap by proving that both the \esl\ and \esfl\ are NP-hard. Additionally, we prove that both problems admit a polynomial-time approximation scheme (PTAS), by demonstrating that approximation algorithms for the \est\ can be adapted to the \esl\ and \esfl\ with appropriate modifications. Specifically, we show $\esfl\leq_{\text{PTAS}}\est$ and $\esl\leq_{\text{PTAS}}\est$, i.e., provide a PTAS reduction to the \est.
\end{abstract}

\section{Introduction}\label{sec:introduction}
The \emph{Euclidean Steiner Tree Problem} (\est) is a fundamental optimization problem that involves finding the shortest tree to interconnect a given set of $n$ points in the Euclidean plane. In contrast to the \emph{Minimum Spanning Tree Problem}, additional intersection points, known as \emph{Steiner points}, may be introduced to further shorten the tree. The problem finds many real-world applications, for example in pipeline layout~\cite{cui2021} or telecommunication systems~\cite{pollak1978}. The first exact algorithm for the \est\ was introduced by Melzak in 1961~\cite{melzak1961}. In 1977, Garey et al.~\cite{garey1977} showed that the \est\ is NP-hard; however, many good approximations exist. Even a simple Minimum Spanning tree has been conjectured by Gilbert and Pollak~\cite{gilbertpollak1968} to provide a $\frac{2}{\sqrt{3}}\approx 1.1547$ approximation. While no counterexample to this conjecture is known, the best proven approximation factor is $1.214$, as shown by Chung and Graham~\cite{chung1985}. Arora~\cite{arora:ptas} and Mitchell~\cite{mitchell1999:ptas} were the first to propose randomized polynomial-time approximation schemes (PTAS), which obtain a $(1+\varepsilon)$-approximation in time $O(n (\log n)^{O(\frac{1}{\varepsilon})})$ and $O(n^{O(\frac{1}{\varepsilon})})$ respectively. Rao and Smith~\cite{raosmith:ptas} later improved this runtime to $O((1/\varepsilon)^{O(\frac{1}{\varepsilon})}n + n\log n)$. Finally, Kisfaludi-Bak et al.~\cite{kisfaludi2021:ptas} reduced the dependence on $\varepsilon$ to $O(2^{O(\frac{1}{\varepsilon})}n + \text{poly}(\frac{1}{\varepsilon})n\log n)$.

In this work, we consider two variants of the \est\ which include an additional straight-line $\gamma$ with zero-cost that has to be incorporated into the tree. In the Euclidean Steiner Line Problem (\esl), the optimal location of this line has to be determined. Conversely, in the Euclidean Steiner fixed Line Problem (\esfl), the line $\gamma$ is provided as input along with the set of terminal points. Here, the line can be regarded as an additional terminal, requiring it to be connected to the tree at least at one point. In practical applications, this line can be thought of an internet cable that is already in place, and now needs to be \emph{last-mile} connected to individual homes. The \esl\ could find applications in determining the optimal location of a straight highway, minimizing the length of additional roads that have to be built to interconnect a set of cities.
Both problems were introduced by Holby~\cite{holby2017}, who proposed a heuristic for the \esfl\ without proving any theoretical approximation guarantees. The first approximation algorithm was presented by Li et  al.~\cite{li2020:line} who demonstrated that similar to the \est , a simple minimum spanning tree constructed on the graph defined by the Euclidean distances between all terminal-terminal and terminal-line pairs achieves a $1.214$ approximation for the \esfl. 
They further proved a conjecture by Holby~\cite{holby2017}, stating that there is always an optimal solution to the \esl, where the line passes through at least two terminal points. Hence, any approximation algorithm for the \esfl\ carries over to the \esl, simply by applying it to each of the $\binom{n}{2}$ possible placements of the line and selecting the best resulting solution. This allows us to focus on the \esfl\ when designing our PTAS, as results for the \esl\ follow directly.

Formally, we define the Euclidean Steiner fixed Line Problem as follows:

\begin{definition}[\esfl ]\label{def:esfl}\

\textbf{Input:} Finite set of terminals $R = \{v_i \in \mathbb{R}^2 \mid i \in \{1, \dots, n\}\}$ and a line $\gamma = \{(x, y) \in \mathbb{R}^2 \mid a\cdot x + b\cdot y = c\}$.

\textbf{Feasible solutions:} Tree $T = (V, E)$ such that $V = (R  \cup S \cup \{\gamma\})$, where $S = \{s_i \in \mathbb{R}^2 \mid i \in \{1, \dots, m\}\}$ is a finite set of points, called \textbf{Steiner points}.

\textbf{Objective:}
$\text{minimize}\ \sum_{(v, w) \in E}d(v, w)$
\end{definition}

Note that in the above definition, $\gamma$ is treated as a node in the tree $T$ and $d$ represents the Euclidean distance between point-point and point-line pairs. The Euclidean Steiner Tree Problem (EST) can be obtained from this definition by omitting $\gamma$. The Euclidean Steiner Line Problem (ESL) is defined similarly to Definition \ref{def:esfl}, except that $\gamma$ is part of the solution rather than part of the input. We refer to the additional points in $V\setminus (R\cup \{\gamma\})$ as \emph{Steiner points}.

\paragraph*{Our contributions}

In this work, we prove a conjecture by Holby~\cite{holby2017}, establishing NP-hardness for both problems through a reduction from a special case of the EST that is known to be NP-hard. Furthermore, we resolve an open question posed by Li et al.~\cite{li2020:line}, by proving that the \esfl, and thus also the \esl, admits a PTAS. Specifically, we show $\esfl\leq_{\text{PTAS}}\est$, i.e., provide a PTAS reduction from the \esfl\ to the \est.

In the following section, we prove NP-hardness for both the \esl\ and the \esfl:

\begin{restatable}{theorem}
{nphardness}\label{thm:nphard} 
    The \esfl\ and the \esl\ are NP-hard.
\end{restatable}

In Section \ref{sec:ptas} we present a PTAS for the \esfl. We begin by establishing a lower bound on the optimal solution value in the \esfl. Next, we introduce a straightforward method to transform an \esfl-instance into an EST instance that can then be solved using a PTAS for the \est. We demonstrate that from the resulting solution, we can construct a solution to the original \esfl-instance that preserves certain approximation guarantees:

\begin{restatable}{theorem}
{ptas}\label{thm:ptas} 
    There exists a randomized polynomial-time approximation scheme for the \esfl\ with runtime $O\left(2^{O(\frac{1}{\varepsilon})}\cdot \frac{n}{\varepsilon} + \textnormal{poly}\left(\frac{1}{\varepsilon}\right) \frac{n}{\varepsilon} \log\left(\frac{n}{\varepsilon}\right) + \left(\frac{n}{\varepsilon}\right)^3\right)$.
\end{restatable}

From the observation of Li et al.~\cite{li2020:line} (Lemma \ref{lemma:li}), it then immediately follows that:
\begin{restatable}{corollary}
{ptascor}\label{cor:ptas} 
    There exists a randomized polynomial-time approximation scheme for the \esl\ with runtime $O\left(n^2 \cdot \left[ 2^{O(\frac{1}{\varepsilon})}\cdot \frac{n}{\varepsilon} + \textnormal{poly}\left(\frac{1}{\varepsilon}\right) \frac{n}{\varepsilon} \log\left(\frac{n}{\varepsilon}\right) + \left(\frac{n}{\varepsilon}\right)^3 \right]\right)$.
\end{restatable}

\paragraph*{Related problems}
Over the past five years, significant progress has been made on problems closely related to the \esl\ and \esfl, with several constant-factor approximation algorithms being proposed.
Li et al.~\cite{li2022:rectilinear} studied variants of the \esfl\ and \esl\ under the rectilinear metric instead of the Euclidean metric. They presented $1.5$-approximation algorithms for both problems, with respective runtimes in $O(n \log n)$ and $O(n^3 \log n)$. In another work, Li et al.~\cite{li2024:segments} considered a generalization of the \esfl\ and \esl, where the terminals are line segments instead of points and presented a $1.214$-approximation algorithms for both problems, again with respective runtimes in $O(n \log n)$ and $O(n^3 \log n)$. Althaus et al.~\cite{althaus2020} studied a similar generalization of the \esfl\ that permits multiple line segments of arbitrary lengths. They developed an exact algorithm based on \textsc{GeoSteiner}~\cite{geosteiner96}, a fast software package for solving the \est. 

Li et al.~\cite{li2021:minsteiner} also analyzed a variant of the \esfl\ with a modified objective function. In this variant, a constant $K$ is provided along with the terminal points and the line $\gamma$, restricting the sum of the lengths of all edges not lying on the line to a maximum of $K$. Furthermore, all Steiner points must be placed on the line, and the objective is to minimize the number of Steiner points. They proposed a $3$-approximation for this problem with a runtime in $O(n^2 \log n)$. Liu~\cite{liu2024:minsteiner} considered a related variant where Steiner points are not constrained to the line and instead of the total sum, each edge not lying on the line is constrained to a maximum length of $K$. They presented a $4$-approximation with runtime $O(n^3)$.

Another related problem setting is the \emph{line-restricted Steiner problem}. Here, the input consists of a set of terminal points and a line $\gamma$, with the restriction that all Steiner points must lie on the line. For the design of exact algorithms, it is useful to consider the variant, where only up to $k$ Steiner points may be placed, namely the \emph{$k$-line-restricted Steiner problem}.
Bose et al.~\cite{bose2020} considered a variant of this problem, where $\gamma$ does not shorten the edge lengths, meaning the length of each edge is simply its Euclidean distance. They presented an exact algorithm for $k>1$ with runtime $O(n^k)$ as well as an algorithm for $k=1$ with runtime $O(n \log n)$.
Li et al.~\cite{li2023:bottleneck} considered a variation of this, the \emph{line-constrained bottleneck $k$-Steiner tree problem}, where edges on the line have zero-cost and instead of the total length of the tree, the length of the largest edge has to be minimized. They presented an exact algorithm with runtime $O(n\log n + n^k\cdot f(k))$, where $f(k)$ is a function only dependent on $k$.

\section{Preliminaries}\label{sec:preliminaries}

We begin with some basic definitions for the \est, proposed by Gilbert and Pollak~\cite{gilbertpollak1968}.

The \emph{topology} of a tree refers to its combinatorial structure of nodes and edges, independent of any geometric representation, i.e., the position of the nodes. A \emph{Steiner topology} is the topological representation of a feasible solution for the \est, where the terminals have at most degree $3$, and the Steiner points have exactly degree $3$ (\emph{degree condition}).
A \emph{Steiner tree} is a geometric realization of a Steiner topology, where no two incident edges form an angle of less than $\ang{120}$, and the three incident edges of each Steiner point meet at exactly $\ang{120}$ (\emph{angle condition}).  
We extend those definitions to the \esfl\ and the \esl, building on the work of Li et al.~\cite{li2020:line}. In these problem settings, the degree condition for points in a Steiner topology remains unchanged, but the line $\gamma$ can have an arbitrary number of connections. For a Steiner tree, the angle conditions for the points are preserved, but edges incident to $\gamma$ must be perpendicular to it.

Based on the degree condition, Gilbert and Pollak~\cite{gilbertpollak1968} demonstrated that a Steiner tree contains at most $n-2$ Steiner points. If a Steiner tree has exactly $n-2$ Steiner points, it is referred to as a \emph{Full Steiner Tree} (FST). We consider two useful properties for the \est, proven by Gilbert and Pollak~\cite{gilbertpollak1968}, and extend them to the \esfl\ and \esl. The first property states that any optimal Steiner tree can be decomposed into a union of FSTs of subsets of the terminals: Every Steiner tree $T$ that is not an FST itself can be decomposed as follows: Replace every terminal $x \in (R \cup \{\gamma\})$ with $k=\delta_T(x) \geq 2$, where $\delta_T(x)$ represents the degree of $x$ in $T$, by $k$ copies of the terminal, each at the same geometric position. Assign each copy as the endpoint of one of the original edges of $x$. This divides $T$ into $k$ smaller trees, where each copy of $x$ now has degree $1$. By iteratively applying this process, we obtain a decomposition of $T$ into trees where each terminal has degree $1$, i.e., into FSTs.

Another well-known property for the \est\ is the \emph{wedge-property}. It states that in an optimal solution, any open wedge-shaped region with an interior angle of $\ang{120}$ or greater that contains no terminals will also contain no Steiner points. Similarly, we will show for the \esfl\ that in an optimal solution, any open wedge-shaped region $W$ with an interior angle of $\ang{120}$ or more, which contains no terminal points and does not intersect $\gamma$, also cannot contain any Steiner points.

\begin{figure}[htb]
    \centering
\scalebox{0.9}{
\begin{tikzpicture}[scale=0.7]
\fill[gray!30, opacity=0.3] (0, 0) -- (-7, 4.04145188432738) -- (7, 4.04145188432738) -- cycle;
    \draw[->, thick] (0, 0) -- (-7, 4.04145188432738);
\draw[->, thick] (0, 0) -- (7, 4.04145188432738);

\draw[-, thick] (-9, -2) -- (9, 0);
\coordinate (A) at (-0.6, 0.34641);
\coordinate (B) at (0, 0);
\coordinate (C) at (0.6, 0.34641);

\fill (2, 0.5) circle (3pt);
\fill (-4, 1.5) circle (3pt);
\fill (3, -1.5) circle (3pt);
\fill (0, -0.4) circle (3pt);
\fill (5, 2.5) circle (3pt);
\fill (-6, 3) circle (3pt);

\pic[draw, angle eccentricity=1.2, angle radius=0.9cm] {angle=C--B--A};
\node[] at (-0,0.7) {$\ang{120}$};
\node[] at (-0,3) {$W$};
\node[] at (-1,-1.6) {$\gamma$};

\fill (2, 3) circle (2pt); \node[] at (2.3, 3.5) {$s$};

\draw[->, dashed, opacity=0.5] (2, 3) -- (1., 2.42264);

\draw[->, dashed, opacity=0.5] (2, 3) -- (3., 2.42264);


, 3.5773);
\draw[->, dashed, opacity=0.5] (2, 3) -- (2.,4);

\end{tikzpicture}
}
    \caption{Illustration of a wedge $W$ (shaded in gray) that does not contain any terminals and has no intersection with $\gamma$. Let $s$ be the Steiner point in $W$ with the largest $y$-coordinate. As at least one edge of $s$ faces upwards and cannot leave the wedge, we arrive at a contradiction, proving that in any optimal solution $W$ does not contain any Steiner points.}
    \label{fig:wedge}
\end{figure}
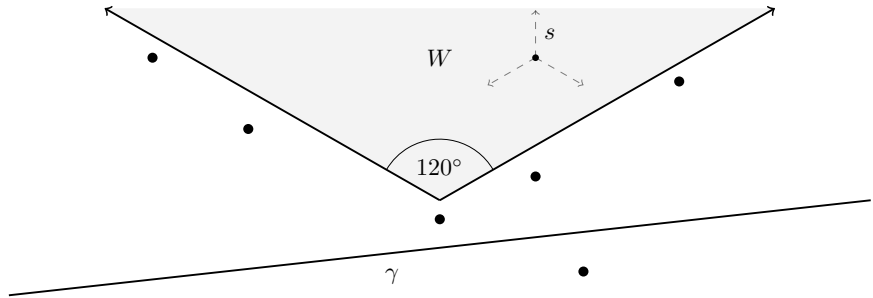

Consider such a region $W$, as illustrated in Figure \ref{fig:wedge}. Without loss of generality, we can assume that $W$ opens symmetrically upwards, as otherwise it would be possible to rotate the instance to satisfy this property. Suppose that $W$ contains at least one Steiner point. Let $s$ be the Steiner point in $W$ with the largest $y$-coordinate. Due to the angle-property, the edges of $s$ must meet at an angle of \ang{120}, thus one of these edges must leave $s$ in a direction within $\pm \ang{60}$ of the positive $y$-axis. This edge cannot leave $W$, therefore its endpoint cannot be a terminal. Consequently, its endpoint must be a another Steiner point in $W$ with a larger $y$-coordinate, resulting in a contradiction.

Another useful result, specifically for the \esl, was conjectured by Holby~\cite{holby2017} and later proven by Li et al.~\cite{li2020:line}:
\begin{lemma}[Li et al.~\cite{li2020:line}]\label{lemma:li}
   Given an instance $I = \{v_1, \dots, v_n \}$ for the \esl, there exists an optimal solution in which $\gamma$ passes through at least two terminal points.
\end{lemma}
In our PTAS reduction, we will make use of a result by Kisfaludi-Bak, Nederlof and Węgrzyc~\cite{kisfaludi2021:ptas}. Their algorithm is currently the fastest known PTAS for the \est.
\begin{theorem}[Kisfaludi-Bak et al.~\cite{kisfaludi2021:ptas}] \label{thm:kisfaludi}
    There exists a randomized polynomial-time approximation scheme for the \est\ that runs in $O\left(2^{O(\frac{1}{\varepsilon})} n + \text{poly}\left(\frac{1}{\varepsilon}\right) n \log(n) \right)$ time.
\end{theorem}
For the proof of Theorem \ref{thm:nphard} we need the following technicality. It can be shown by a simple case distinction that is included in the Appendix.
\begin{restatable}{observation}
{observationCases}\label{observation:abs} 
    For any $a \in \mathbb{R}$ it holds that $|5a + 3| + |3a + 5| > \frac{5}{2}\sqrt{a^2+1}.$
\end{restatable}

\section{NP-hardness}\label{sec:nphardness}
Rubinstein et al.~\cite{rubinstein1997} showed that the \est\ remains NP-hard when all terminals are restricted to two parallel lines, a variant we refer to as \textsc{PALIMEST} (\textbf{Pa}rallel \textbf{Li}ne \textbf{M}inimal \textbf{E}uclidean \textbf{S}teiner \textbf{T}ree). We reduce \textsc{PALIMEST} to the \esfl\ and the \esl\ to show the following result:
\nphardness*
\begin{proof}
We begin by proving the NP-hardness of the \esfl\ and then demonstrate how this proof can be adapted for the \esl. For the proof, we consider the respective decision variants of these problems, where an additional parameter $k$ is provided, and it must be determined whether a Steiner tree of length $\leq k$ exists.

\paragraph*{NP-hardness of the \esfl}

We show $\textsc{PALIMEST} \leq_{p} \esfl$:
Given an instance $I = (R \subset \mathbb{R}^2, k)$ for the decision variant of $\textsc{PALIMEST}$, we may assume without loss of generality that the two lines are horizontal, all points have non-negative coordinates, and $(0, 0) \in R$. For an instance $I$ of $\textsc{PALIMEST}$, we construct an instance $I'$ of the \esfl\ as follows:

Let $M$ be the length of a minimum spanning tree (MST) of $R$. Define $\gamma := \{(x, y) \in \mathbb{R}^2 \mid x + y = -2M\}$, which is a line with slope $-1$ passing through the point $(-M, -M)$. Using this, we construct the \esfl-instance $I' = (R, \gamma, k' = k + \sqrt{2}M)$. We prove that there is a solution for $I$ with cost at most $k$ if and only if there is a solution for $I'$ with cost at most $k'=k+\sqrt{2}M$. The high-level idea behind this construction is that the distance between the line and the terminal points is sufficiently large to ensure that the line only has a single connection. We will further argue that this connection is precisely the edge $\{\gamma, (0,0)\}$ with length $\sqrt{2}M$. It then follows that $OPT_{I'} = OPT_I + \sqrt{2}M$. An illustration of the construction is depicted in Figure \ref{fig:hardness:overview}.

\begin{figure}[htb]
    \centering
    \begin{subfigure}[t]{0.48\textwidth}
        \centering
        \resizebox{\textwidth}{!}{\definecolor{lgray}{RGB}{210,210,210}
\definecolor{wgreen}{RGB}{52, 189, 0}
\begin{tikzpicture}[scale=1.15]

\draw[line width=1,wgreen] (0.33747825766307793, 0.6192359719375004) -- (0.5, 0.75);
\draw[line width=1,wgreen] (0.4165182635630146, 0.10984178357249963) -- (0.33747825766307793, 0.6192359719375004);
\draw[line width=1,wgreen] (0.7, 0.0) -- (0.4165182635630146, 0.10984178357249963);
\draw[line width=1,wgreen] (0.28, 0.0) -- (0.4165182635630146, 0.10984178357249963);
\draw[line width=1,wgreen] (0.0, 0.75) -- (0.33747825766307793, 0.6192359719375004);
\draw[line width=1,wgreen] (1.393915149962827, 0.6111630223085582) -- (1.2, 0.75);
\draw[line width=1,wgreen] (1.7, 0.75) -- (1.393915149962827, 0.6111630223085582);
\draw[line width=1,wgreen] (1.3448678800297387, 0.11106958215315327) -- (1.393915149962827, 0.6111630223085582);
\draw[line width=1,wgreen] (1.5, 0.0) -- (1.3448678800297387, 0.11106958215315327);
\draw[line width=1,wgreen] (1.1, 0.0) -- (1.3448678800297387, 0.11106958215315327);
\draw[line width=1,wgreen] (1.7, 0.75) -- (1.9, 0.75);
\draw[line width=1,wgreen] (0.0, 0.0) -- (0.28, 0.0);
\draw[line width=1,wgreen] (1.5, 0.0) -- (1.8, 0.0);
\draw[line width=1,wgreen] (0.7, 0.0) -- (1.1, 0.0);
\fill[wgreen] (0.33747825766307793, 0.6192359719375004) circle (2pt);
\fill[wgreen] (0.4165182635630146, 0.10984178357249963) circle (2pt);
\fill[wgreen] (1.393915149962827, 0.6111630223085582) circle (2pt);
\fill[wgreen] (1.3448678800297387, 0.11106958215315327) circle (2pt);

    \foreach \x in {0, 0.28, 0.7, 1.1, 1.8, 1.5} {
        \fill (\x, 0) circle (2pt);
    }

    \foreach \x in {0, 0.5, 1.7, 1.2, 1.9} {
        \fill (\x, 0.75) circle (2pt);
    }


\draw[->, line width=0.1mm, opacity=0.6] (-1.5, 0) -- (3.5, 0) node[right] {$x$};
    \draw[->, opacity=0.6] (0, -2) -- (0, 2) node[above] {$y$};

\end{tikzpicture}}
        \caption{A \textsc{PALIMEST}-instance $I$ with an optimal solution. Without loss of generality, the two parallel lines are horizontal, all coordinates are non-negative and $I$ includes the point $(0 ,0)$.}\label{fig:hardness:overview:palimest}
    \end{subfigure}
    \hfill
    \begin{subfigure}[t]{0.48\textwidth}
        \centering
        \resizebox{\textwidth}{!}{\definecolor{lgray}{RGB}{210,210,210}
\definecolor{wgreen}{RGB}{52, 189, 0}
\begin{tikzpicture}[scale=0.65]

    


\draw[line width=1,wgreen] (0.33747825766307793, 0.6192359719375004) -- (0.5, 0.75);
\draw[line width=1,wgreen] (0.4165182635630146, 0.10984178357249963) -- (0.33747825766307793, 0.6192359719375004);
\draw[line width=1,wgreen] (0.7, 0.0) -- (0.4165182635630146, 0.10984178357249963);
\draw[line width=1,wgreen] (0.28, 0.0) -- (0.4165182635630146, 0.10984178357249963);
\draw[line width=1,wgreen] (0.0, 0.75) -- (0.33747825766307793, 0.6192359719375004);
\draw[line width=1,wgreen] (1.393915149962827, 0.6111630223085582) -- (1.2, 0.75);
\draw[line width=1,wgreen] (1.7, 0.75) -- (1.393915149962827, 0.6111630223085582);
\draw[line width=1,wgreen] (1.3448678800297387, 0.11106958215315327) -- (1.393915149962827, 0.6111630223085582);
\draw[line width=1,wgreen] (1.5, 0.0) -- (1.3448678800297387, 0.11106958215315327);
\draw[line width=1,wgreen] (1.1, 0.0) -- (1.3448678800297387, 0.11106958215315327);
\draw[line width=1] (1.7, 0.75) -- (1.9, 0.75);
\draw[line width=1] (0.0, 0.0) -- (0.28, 0.0);
\draw[line width=1] (1.5, 0.0) -- (1.8, 0.0);
\draw[line width=1] (0.7, 0.0) -- (1.1, 0.0);
\fill[wgreen] (0.33747825766307793, 0.6192359719375004) circle (2pt);
\fill[wgreen] (0.4165182635630146, 0.10984178357249963) circle (2pt);
\fill[wgreen] (1.393915149962827, 0.6111630223085582) circle (2pt);
\fill[wgreen] (1.3448678800297387, 0.11106958215315327) circle (2pt);

    \foreach \x in {0, 0.28, 0.7, 1.1, 1.8, 1.5} {
        \fill (\x, 0) circle (2pt);
    }

    \foreach \x in {0, 0.5, 1.7, 1.2, 1.9} {
        \fill (\x, 0.75) circle (2pt);
    }


    
    \draw[thick, red] (2, -5) -- (-5, 2);
    \draw[thick, black] (0, 0) -- (-1.5, -1.5);
    \node[below left] at (-1, -2) {$\gamma: y = -x - 2M$};

    \coordinate (p) at (2*6/4, -4*6/4);
    \coordinate (q) at (-4*6/4, 2*6/4);

    
    
    \draw[->, line width=0.1mm, opacity=0.6] (-4.5, 0) -- (3.5, 0) node[right] {$x$};
    \draw[->, opacity=0.6] (0, -4) -- (0, 2) node[above] {$y$};

    \coordinate (O) at (-0.6, 0.2);

    \def\angleA{105}
    \def\angleB{225}

    \def\radius{2}
    \def\radiuss{1.3}

    \coordinate (A) at ($ (O) + (\angleA:\radius) $);
    \coordinate (B) at ($ (O) + (\angleB:\radius) $);

    \coordinate (R) at ($ (O) + (\angleA:\radiuss) $);
    \coordinate (T) at ($ (O) + (\angleB:\radiuss) $);

\end{tikzpicture}}
        \caption{An \esfl-instance $I'$ obtained by adding the line $\gamma: y=-x-2M$ to $I$. The optimal solution contains the edge $\{\gamma, (0,0)\}$, while the rest of the solution remains identical to an optimal solution for $I$. }\label{fig:hardness:overview:line}
    \end{subfigure}
    \caption{Illustration of the construction we use for reducing \textsc{PALIMEST} to the \esfl. Terminals are depicted in black, Steiner points in green.}
    \label{fig:hardness:overview}
\end{figure}
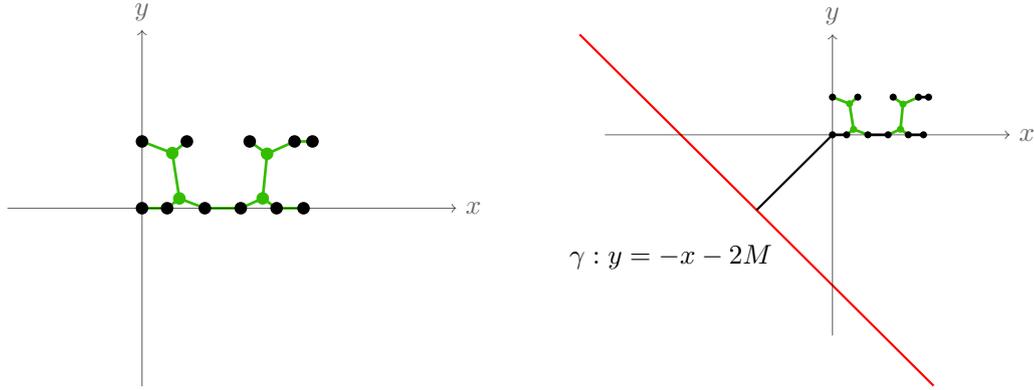
To show $OPT_{I'} \leq OPT_I + \sqrt{2}M$, let $T_I^*$ be a solution for $I$ with cost $OPT_I$. Then $T_{I'} = T_I^* \cup \{\{\gamma, (0, 0)\}\}$ is a solution for $I'$ with cost $OPT_I + \sqrt{2}M$. Thus, $OPT_{I'} \leq OPT_I + \sqrt{2}M$.

To show $OPT_{I'} \geq OPT_I + \sqrt{2}M$, it suffices to prove that in any optimal solution of $I'$, the line $\gamma$ must be directly connected to the point $(0, 0)$:
First, we show that the line $\gamma$ has exactly one incident edge with length at most $\sqrt{2}M$. Suppose $\gamma$ has multiple incident edges. Then, $\gamma$ would belong to at least two different full Steiner trees (FSTs) included in the solution. However, the weight of each FST is at least $\sqrt{2}M$, since the closest terminal point to the line is at a distance of $\sqrt{2}M$. This would imply $OPT_{I'} \geq 2\sqrt{2}M$, which contradicts $OPT_I + \sqrt{2}M \geq OPT_{I'}$ since $OPT_I < \sqrt{2}M$. Therefore, $\gamma$ has only one incident edge $e = \{\gamma, A\}$, where $A$ is the adjacent point of $\gamma$. Further note that $c(e) \leq \sqrt{2}M$, as $e$ could otherwise be easily replaced with the edge $\{\gamma, (0, 0)\}$ with $c(\{\gamma, (0, 0)\})=\sqrt{2}M$, contradicting the optimality of the solution.

Next, we show that in any optimal solution of $I'$, the line $\gamma$ must be directly connected to the origin terminal, i.e., $A=(0,0)$. Suppose $A \neq (0, 0)$. Then, $A$ must have either a negative $x$-coordinate, a negative $y$-coordinate, or both; otherwise, the edge length would exceed $\sqrt{2}M$. Without loss of generality, suppose $A$ has a negative $x$-coordinate (if not, the entire instance can be reflected about the axes' bisector). Thus, $A$ cannot be a terminal point and must be a Steiner point with three edges meeting at $120^\circ$ angles. Let $v_1$ be the vector representing the direction of the edge from $A$ to $\gamma$, and let $v_2$ represent the direction of the next edge in clockwise direction, pointing in positive $y$-direction and negative $x$-direction. Specifically, $v_1 = (-1, -1)$ and $v_2 = (-1, \tan 75^\circ)$. Define $H = \{(x, y) \in \mathbb{R}^2 \mid x < A_x\}$ as the set of points with $x$-coordinate smaller than $A_x$ (see Figure \ref{fig:hardness:esfl}), and let $C \neq A$ be the point in $H$ with the smallest $x$-coordinate. Since $H$ contains no terminal points, $C$ cannot be a terminal. Therefore, $C$ must be a Steiner point. Due to the $120^\circ$ angle property, $C$ must have a neighbor within $H$ with a smaller $x$-coordinate. However, this is impossible since there are no terminal points in~$H$ and no Steiner points with a smaller $x$-coordinate than $C$. Since $\gamma$ has only degree one and already shares an edge with $A$, it cannot also be adjacent to $C$. Thus, such a point $C$ cannot exist, implying that there are no Steiner points in $H$. In particular, there are no points on the ray $\{A + r \cdot v_2 \mid r > 0\}$, so $A$ cannot have a neighbor in that direction. We arrive at a contradiction, showing that $A = (0, 0)$.

To conclude, let $T_{I'}^*$ be an optimal solution for $I'$. Since the only incident edge of $\gamma$ is $\{\gamma, (0, 0)\}$, the tree $T_I = T_{I'}^*[V(T_{I'}^*)\setminus \{\gamma\}]$ is a solution to $I$ with cost $OPT_{I'} - \sqrt{2}M$. It follows that $OPT_{I'} \geq OPT_I + \sqrt{2}M$.

Therefore, there is a solution for $I$ with cost at most $k$ if and only if there is a solution for $I'$ with cost at most $k' = k + \sqrt{2}M$ and we can conclude that the \esfl\ is NP-hard.

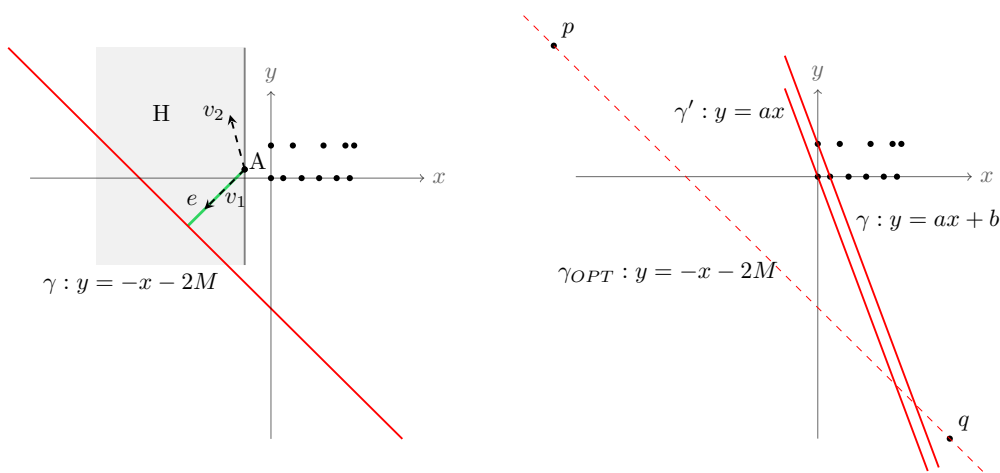
\begin{figure}
    \centering
    \begin{subfigure}[t]{0.48\textwidth}
        \centering
        \resizebox{1\textwidth}{!}{
        \definecolor{lgray}{RGB}{210,210,210}
\definecolor{wgreen}{RGB}{47, 212, 96}
\begin{tikzpicture}[scale=0.65]

    \draw[white] (2.5*6/4, -4.5*6/4) -- (-4.5*6/4, 2.5*6/4);
    
    
    \fill[lgray, opacity=0.3] (-4,3) rectangle (-0.6,-2);

    \draw[gray, thick] (-0.6, 3) -- (-0.6, -2);

    \foreach \x in {0, 0.28, 0.7, 1.1, 1.8, 1.5} {
        \fill (\x, 0) circle (2pt);
    }

    \foreach \x in {0, 0.5, 1.7, 1.2, 1.9} {
        \fill (\x, 0.75) circle (2pt);
    }

    \draw[very thick, wgreen] (-1.9, -1.1) -- (-0.6, 0.2);\node[] at (-1.8, -0.5)  {$e$};


    \fill (-0.6, 0.2) circle (2pt); \node[] at (-0.3, 0.4) {A};


    
    \draw[thick, red] (2*6/4, -4*6/4) -- (-4*6/4, 2*6/4);
    \node[below left] at (-1, -2) {$\gamma: y = -x - 2M$};

    \coordinate (p) at (2*6/4, -4*6/4);
    \coordinate (q) at (-4*6/4, 2*6/4);

    
    
    \draw[->,opacity=0.6] (-5.5, 0) -- (3.5, 0) node[right] {$x$};
    \draw[->,opacity=0.6] (0, -6) -- (0, 2) node[above] {$y$};

    \coordinate (O) at (-0.6, 0.2);

    \def\angleA{105}
    \def\angleB{225}

    \def\radius{2}
    \def\radiuss{1.3}

    \coordinate (A) at ($ (O) + (\angleA:\radius) $);
    \coordinate (B) at ($ (O) + (\angleB:\radius) $);

    \coordinate (R) at ($ (O) + (\angleA:\radiuss) $);
    \coordinate (T) at ($ (O) + (\angleB:\radiuss) $);

    \draw[black, thick, dashed, -stealth] (O) -- (T);\node[] at (-0.8, -0.5) {$v_1$};
    \draw[black, thick, dashed, -stealth] (O) -- (R) node[near end, above left] {$v_2$};

    \fill (O) circle (2pt);
    \node at (-2.5, 1.5) {H};

\end{tikzpicture}}
        \caption{In the \esfl-instance $I'$, $\gamma$ cannot have a connection to a Steiner point $A$ with a negative coordinate due to the angle property of $A$. Otherwise, there would need to exist a Steiner point in the direction of $v_2$, but since there are no Steiner points in $H$, we arrive at a contradiction.}\label{fig:hardness:esfl}
    \end{subfigure}
    \hfill
    \begin{subfigure}[t]{0.48\textwidth}
        \centering
        \resizebox{1\textwidth}{!}{\definecolor{lgray}{RGB}{210,210,210}
\definecolor{wgreen}{RGB}{47, 212, 96}
\begin{tikzpicture}[scale=0.65]


    \foreach \x in {0, 0.28, 0.7, 1.1, 1.8, 1.5} {
        \fill (\x, 0) circle (2pt);
    }

    \foreach \x in {0, 0.5, 1.7, 1.2, 1.9} {
        \fill (\x, 0.75) circle (2pt);
    }


    
    \draw[thick, red] (-0.75, 2.77) -- (2.75, -6.66);
    \draw[thick, red] (-0.75, 2.02) -- (2.5, 0.6742423636363637-7.416);
    \node[] at (2.5, -1) {$\gamma: y=ax+b$};
    \node[] at (-2, 1.5) {$\gamma': y=ax$};
    
    \node[] at (-3.4, -2.2) {$\gamma_{OPT}: y = -x - 2M$};


    \draw[->,opacity=0.6] (-5.5, 0) -- (3.5, 0) node[right] {$x$};
    \draw[->,opacity=0.6] (0, -6) -- (0, 2) node[above] {$y$};

    \coordinate (p) at (2*6/4, -4*6/4);
    \coordinate (q) at (-4*6/4, 2*6/4);
        

    \fill (p) circle (2pt) node[above right] {$q$};
    \fill (q) circle (2pt) node[above right] {$p$};

    \draw[red, dashed] (2.5*6/4, -4.5*6/4) -- (-4.5*6/4, 2.5*6/4);








\end{tikzpicture}}
        \caption{Extension of the reduction to the \esl\ by adding the points $p$ and $q$ that enforce $\gamma_{OPT}$ to be \mbox{$y=-x-2M$} (dashed line). Assuming $\gamma$ does not pass through $p$ or $q$, it could be shifted down to the origin as the sum of distances from $p$ and $q$ to $\gamma'$ can only decrease. We show this sum exceeds the upper bound, leading to a contradiction.}\label{fig:hardness:esl}
    \end{subfigure}
    \caption{Illustration of the NP-hardness proofs presented in Theorem \ref{thm:nphard} for the \esfl\ and \esl, respectively.}
    \label{fig:hardness}
\end{figure}
\paragraph*{Extension to the \esl}
We now demonstrate that the reduction can be adapted to prove NP-hardness of the \esl. Instead of adding a fixed line $\gamma$, we introduce two terminals, which enforce that in any optimal solution, the placed line is $y = -x - 2M$. Specifically, we add the points $p = (-5M, 3M)$ and $q = (3M, -5M)$, constructing an instance $I' = (R\cup \{p, q\}, k' = k + \sqrt{2}M)$ for the ESL. Due to the remote positions of $p$ and $q$, the line passing through any other pair of points than $p$ and $q$ would result in the solution length exceeding the upper bound.

Using the same argument as for the \esfl, it holds that $OPT_{I'} \leq OPT_I + \sqrt{2}M$. Observe that to show $OPT_{I'} \geq OPT_I + \sqrt{2}M$, it suffices to demonstrate that $\gamma = \{(x, y) \in \mathbb{R}^2 \mid y = -x - 2M\}$, as the rest immediately follows from the proof for the \esfl.

Suppose, for contradiction, that $\gamma \neq \{(x, y) \in \mathbb{R}^2 \mid y = -x - 2M\}$ in an optimal solution, i.e., the line does not pass through $p$ and $q$. We assume that $p$ and $q$ would not be connected when removing $\gamma$, as otherwise the solution length would be at least $d(p, q)=\sqrt{128} M$. This would lead to a contradiction, as in this case $\sqrt{128} M = d(p, q) \leq OPT_{I'}$, but $OPT_{I'} \leq (1 + \sqrt{2})M$. Further, we can assume that the placed line is not vertical; otherwise, $d(p, \gamma) + d(q, \gamma)$ would be at least $8M$, because the horizontal distance of $p$ and $q$ is exactly $8M$. Since $d(p, \gamma) + d(q, \gamma) \leq OPT_{I'} \leq (1+\sqrt{2}M)$, this would lead to a contradiction. Therefore, $\gamma$ can be written in the form $y = ax + b$.

Next, we observe that the line must pass through the second and fourth quadrant and include some points with non-negative coordinates (in both dimensions). If $\gamma$ would not include points with non-negative coordinates, then by Lemma \ref{lemma:li}, it would need to pass through $p$ and $q$, as these are the only two terminals with a negative coordinate. This would contradict our original assumption. If $\gamma$ would not pass through the second or fourth quadrant, then either $d(p, \gamma)$ or $d(q, \gamma)$ would be at least $3M$, yielding another contradiction, as $OPT_{I'} \leq (1 + \sqrt{2})M$. Hence, we can assume that $b \geq 0$.

The distance between a line $y=ax+b$ and a point $(x_0, y_0)$ is given by $\frac{|a\cdot x_0-y_0+b|}{\sqrt{a^2+1}}$. 
To arrive at a contradiction, we now show that $d(p, \gamma) + d(q, \gamma)$ exceeds $\frac{5}{2}M$. We introduce a new line $\gamma':= \{y=ax\}$, which has the same slope as $\gamma$ but a $y$-intercept of $0$ instead of $b$. As $b\geq 0$, $\gamma$ cannot lie below both $p$ and $q$ and shifting down the line from $y$-intercept $b$ to $0$ cannot increase the sum of the distances to $p$ and $q$. It therefore holds that $d(p, \gamma') + d(q, \gamma') \leq d(p, \gamma) + d(q, \gamma)$ (illustrated in Figure \ref{fig:hardness:esl}). 

Using Observation \ref{observation:abs}, we obtain

\begin{align*} 
d(p, \gamma') + d(q, \gamma') &=\quad \frac{|-5M \cdot a - 3M|}{\sqrt{a^2+1}} + \frac{|3M \cdot a + 5M|}{\sqrt{a^2+1}} \\
&= \quad M \left(\frac{|5a + 3| + |3a + 5|}{\sqrt{a^2+1}} \right) \\
&\stackrel{\mathclap{Obs.~\ref{observation:abs}}}{>} \quad \frac{5}{2} M. \\ 
\end{align*} 

    As $d(p, \gamma') + d(q, \gamma')$ is larger than $\frac{5}{2}M$, so is $d(p, \gamma) + d(q, \gamma)$, contradicting $OPT_{I'} \leq (1+\sqrt{2}M)$. We conclude that $\gamma = \{(x, y) \in \mathbb{R}^2 \mid y = -x - 2M\}$. Using the same reasoning as in the proof for the \esfl, we deduce that $OPT_{I'} \geq OPT_I + \sqrt{2}M$, thereby establishing the NP-hardness of the \esl.

\end{proof}

\section{PTAS}\label{sec:ptas}
As outlined in the introduction, this section primarily focuses on the \esfl, as approximations for the \esfl\ extend to the \esl\ with an additional runtime factor of $O(n^2)$.

Let $I=(R \subset \mathbb{R}^2, \gamma := \{(x,y) \in \mathbb{R}^2 \mid a\cdot x + b\cdot y = c\})$ be an instance for the \esfl\ with $n:=|R|$ terminal points and the line $\gamma$. Without loss of generality, in the following we assume $a=c=0$, giving a horizontal line at $y=0$. Furthermore, we assume that all terminal points lie above the line, i.e., have a non-negative $y$-coordinate; otherwise, the subproblems with terminals above and below the line could be solved independently.
In this context, we define $w(P) :=\max_{(x,y), (x', y') \in P}{|x-x'|}$ as the width and $h(P):=\max_{(x,y)\in P}y$ as the height of a point set $P\subset \mathbb{R}^2$ with respect to the line. We further denote $W_I:=w(R)$ as the width of the instance $I$.

\subsection{Lower bound}
To achieve a desirable approximation factor for the \esfl, we must first establish a lower bound on the optimal solution. In contrast to the \est, we cannot simply use the width of the instance as a lower bound, as terminals may lie arbitrarily far apart from each other while still incurring only small costs if they are close to the line. To address this, we will demonstrate that in such cases, the instance can be decomposed by splitting it into subinstances and solving them independently. Using this approach, we derive a lower bound for optimal solutions that depends solely on the width of the instance.
\begin{lemma}[Decomposition]\label{lemmaDecomp}
For an instance $I=(R, \gamma=\{(x,0)\mid x\in \mathbb{R}\})$ of the \esfl\ with the terminal set $R = \{(x_1, y_1), \dots, $$(x_n, y_n)\}$ sorted in non-decreasing order by $x_i$, we can assume without loss of generality,  that for all $i \in \{1, \dots, n-1\}$, it holds that
\[\max_{j \in \{1, \dots, i\}} (x_j + \frac{y_{j}}{\tan(\ang{30})}) \geq \min_{j \in \{i+1, \dots, n\}} (x_j - \frac{y_{j}}{\tan(\ang{30})})\]
as otherwise, the instance could be split at these points, and the subinstances could be solved independently.
\end{lemma}

\begin{proof}
Assume that for some $i \in \{1, \dots, n-1\}$, it holds that $a := \max_{j \in \{1, \dots, i\}} (x_j + \frac{y_{j}}{\tan(\ang{30})}) < \min_{j \in \{i+1, \dots, n\}} (x_j - \frac{y_{j}}{\tan(\ang{30})}) =: b$. Let $j_a \in \{1, \dots, i\}$ and $j_b \in \{i+1, \dots, n\}$ be the respective indices where the maximum and minimum is achieved. Let $\overline{a}$ be the line through the points $(a, 0)$ and $(x_{j_a}, y_{j_a})$, and similarly, let $\overline{b}$ be the line through $(b, 0)$ and $(x_{j_b}, y_{j_b})$ (see Figure \ref{fig:decomposition}).

There can be no terminal $(x_k, y_k)$ with $k \in \{1, \dots, i\}$ lying above $\overline{a}$, as $y_k > y_{j_a} - \tan(\ang{30})(x_k - x_{j_a})$ would imply $x_k + \frac{y_k}{\tan(\ang{30})} > x_{j_a} + \frac{y_{j_a}}{\tan(\ang{30})}$ which contradicts the maximality of $j_a$. Similarly, there is no terminal $(x_l, y_l)$ with $l \in \{i+1, \dots, n\}$ lying above the line $\overline{b}$.

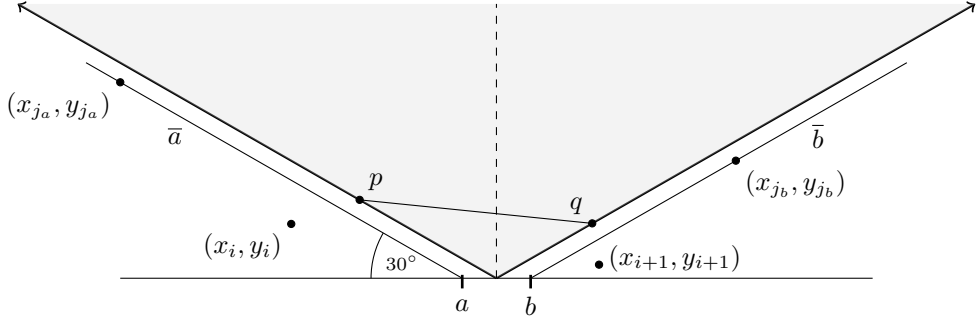
\begin{figure}[htb]
    \centering
    \begin{tikzpicture}[scale=0.9]
\draw (-0.5, 0) -- (-6.0, 3.1754264805429417);

\coordinate (A) at (-5.5, 2.8867513459481287);
\filldraw (A) circle (1.5pt);
\node[below left] at (A) {$(x_{j_a}, y_{j_a})$};

\filldraw (-3, 0.8) circle (1.5pt);
\node[below left] at (-3, 0.8) {$(x_{i}, y_{i})$};
\draw (0.5, 0) -- (6.0, 3.1754264805429417);
\filldraw (3.5, 1.7320508075688772) circle (1.5pt);
\node[below right] at (3.5, 1.7320508075688772) {$(x_{j_b}, y_{j_b})$};
\filldraw (1.5, 0.2) circle (1.5pt);
\node[right] at (1.5, 0.3) {$(x_{i+1}, y_{i+1})$};
\node[] at (-4.7, 2.1) {$\overline{a}$};
\node[] at (4.7, 2.1) {$\overline{b}$};
\node[] at (0.5, -0.4) {$b$};

\coordinate (B) at (-0.5, 0);
\node[] at (-0.5, -0.4) {$a$};
\draw[line width = 1pt] (0.5, 0.05) -- (0.5, -0.15);
\draw[line width = 1pt] (-0.5, 0.05) -- (-0.5, -0.15);
\draw (-5.5, 0) -- (5.5, 0);
\draw[->, thick] (0, 0) -- (-7, 4.04145188432738);
\draw[->, thick] (0, 0) -- (7, 4.04145188432738);
\fill[gray!30, opacity=0.3] (0, 0) -- (-7, 4.04145188432738) -- (7, 4.04145188432738) -- cycle;
\draw[] (-2, 1.1547005383792515) -- (1.4, 0.808290376865476);
\filldraw (1.4, 0.808290376865476) circle (1.5pt);
\node[above left] at (1.4, 0.808290376865476) {$q$};
\filldraw (-2, 1.1547005383792515) circle (1.5pt);
\node[above right] at (-2, 1.1547005383792515) {$p$};
\draw[dashed] (0, 0) -- (0, 4.04145188432738);

\coordinate (C) at (-5, 0);

\pic[draw, angle eccentricity=1.2, angle radius=1.2cm] {angle=A--B--C};
\node[] at (-1.38,0.23) {$\scriptstyle\ang{30}$};

\end{tikzpicture}
    \caption{Illustration of the decomposition in Lemma \ref{lemmaDecomp}. If a symmetric wedge with apex on the line can be drawn without enclosing any terminals, it cannot contain any Steiner points, nor can there be an edge $\{p, q\}$ connecting two points lying on a different border of the wedge, because directly connecting both $p$ and $q$ to the line would shorten the tree. Thus, both sides of the wedge can be solved independently.}
    \label{fig:decomposition}
\end{figure}

We show that in any optimal solution, the terminal sets $R_a := \{(x_1, y_1), \dots, (x_i, y_i)\}$ and $R_b := \{(x_{i+1}, y_{i+1}), \dots, $
$(x_n, y_n)\}$ are only connected via the line. Since $a < b$ and all terminals lie below $\overline{a}$ or $\overline{b}$, there exists a symmetric wedge with apex at $(\frac{a+b}{2}, 0)$, opening upwards with an internal angle of $\ang{120}$ containing no terminals. Using the wedge-property (see Section \ref{sec:preliminaries}), we know that no Steiner points lie inside this wedge.
Suppose there are connections above the line between terminals to the left and right of the wedge. In that case, there would be an edge crossing both borders of the wedge. If $p$ and $q$ are these crossing points, then $\frac{y_p}{\tan(\ang{30})}$ and $\frac{y_q}{\tan(\ang{30})}$ are their respective horizontal distances to the center of the wedge. Since $\frac{1}{\tan(\ang{30})} > 1$, the cost of the solution could be reduced by replacing the edge $\{p, q\}$ with the edges $\{p, \gamma\}$ and $\{q, \gamma\}$.

This contradicts the optimality of the solution, implying that the terminal sets $R_a := \{(x_1, y_1), \dots, (x_{i}, y_{i})\}$ and $R_b := \{(x_{i+1}, y_{i+1}), \dots, (x_n, y_n)\}$ are only connected via the line. Thus, the instances $(R_a, \gamma)$ and $(R_b, \gamma)$ can be solved independently, and their solutions combined to obtain an optimal solution for $I$. The Lemma follows by recursively applying this decomposition to $R_a$ and $R_b$. This proof is illustrated in Figure \ref{fig:decomposition}.
\end{proof}

\begin{figure}[htb]
\centering
    \scalebox{0.9}{
\begin{tikzpicture}

\fill[gray!30, opacity=0.3] (-5.167207792207792, 1.220779220779221) -- (-5.167-2.11, 0) -- (-5.167+2.11, 0) -- cycle;

\coordinate (A) at (-0.8133116883116891, 2.1558441558441555);
\coordinate (B) at (-0.813-3.74, 0);
\coordinate (C) at (-0.813+3.74, 0);
\fill[gray!30, opacity=0.3] (A) -- (B) -- (C) -- cycle;
\pic[draw, angle eccentricity=1.2, angle radius=1.3cm] {angle=C--B--A};
\node[] at (-3.6,0.25) {$\color{black}\scriptstyle\ang{30}$};

\fill[gray!30, opacity=0.3] (2.15, 0.59) -- (1.133, 0) -- (3.175, 0) -- cycle;

\fill[gray!30, opacity=0.3] (4.45, 1.95) -- (1.064, 0) -- (7.84, 0) -- cycle;

\draw [color={rgb, 255:red, 52; green, 189; blue, 0 }, draw opacity=1, line width=1] (1.9176199124709528, 0.0) -- (1.9176199124709528, 0.313521932514813);
\draw [color={rgb, 255:red, 52; green, 189; blue, 0 }, draw opacity=1, line width=1] (1.9176199124709528, 0.313521932514813) -- (1.7970779220779214, 0.38311688311688297);
\draw [color={rgb, 255:red, 52; green, 189; blue, 0 }, draw opacity=1, line width=1] (1.9176199124709528, 0.313521932514813) -- (2.27435064935065, 0.5194805194805197);
\draw [color={rgb, 255:red, 52; green, 189; blue, 0 }, draw opacity=1, line width=1] (4.112137908353681, 0.0) -- (4.112137908353681, 1.303704718055211);
\draw [color={rgb, 255:red, 52; green, 189; blue, 0 }, draw opacity=1, line width=1] (4.112137908353681, 1.303704718055211) -- (3.71590909090909, 1.5324675324675328);
\draw [color={rgb, 255:red, 52; green, 189; blue, 0 }, draw opacity=1, line width=1] (4.112137908353681, 1.303704718055211) -- (4.845779220779221, 1.7272727272727275);
\draw [color={rgb, 255:red, 52; green, 189; blue, 0 }, draw opacity=1, line width=1] (-0.25455170061248933, 0.0) -- (-0.25455170061248933, 0.8653534312652598);
\draw [color={rgb, 255:red, 52; green, 189; blue, 0 }, draw opacity=1, line width=1] (-0.25455170061248933, 0.8653534312652598) -- (-0.8133116883116893, 1.187953660575785);
\draw [color={rgb, 255:red, 52; green, 189; blue, 0 }, draw opacity=1, line width=1] (-0.8133116883116893, 1.187953660575785) -- (-1.1444805194805197, 0.9967532467532472);
\draw [color={rgb, 255:red, 52; green, 189; blue, 0 }, draw opacity=1, line width=1] (-0.8133116883116893, 1.187953660575785) -- (-0.8133116883116891, 2.1558441558441555);
\draw [color={rgb, 255:red, 52; green, 189; blue, 0 }, draw opacity=1, line width=1] (-0.25455170061248933, 0.8653534312652598) -- (-0.043831168831168554, 0.9870129870129869);
\draw [color={rgb, 255:red, 52; green, 189; blue, 0 }, draw opacity=1, line width=1] (-4.650974025974026, 0.0) -- (-4.650974025974026, 0.5584415584415581);
\draw [color={rgb, 255:red, 52; green, 189; blue, 0 }, draw opacity=1, line width=1] (-5.167207792207792, 1.220779220779221) -- (-4.650974025974026, 0.5584415584415581);
\filldraw[color={rgb, 255:red, 52; green, 189; blue, 0 }] (1.9176199124709528, 0.313521932514813) circle (1.5pt);
\filldraw[color={rgb, 255:red, 52; green, 189; blue, 0 }] (4.112137908353681, 1.303704718055211) circle (1.5pt);
\filldraw[color={rgb, 255:red, 52; green, 189; blue, 0 }] (-0.25455170061248933, 0.8653534312652598) circle (1.5pt);
\filldraw[color={rgb, 255:red, 52; green, 189; blue, 0 }] (-0.8133116883116893, 1.187953660575785) circle (1.5pt);
\filldraw[black] (-1.1444805194805197, 0.9967532467532472) circle (2pt);
\filldraw[black] (3.71590909090909, 1.5324675324675328) circle (2pt);
\filldraw[black] (4.845779220779221, 1.7272727272727275) circle (2pt);
\filldraw[black] (-0.043831168831168554, 0.9870129870129869) circle (2pt);
\filldraw[black] (-0.8133116883116891, 2.1558441558441555) circle (2pt);
\filldraw[black] (-4.650974025974026, 0.5584415584415581) circle (2pt);
\filldraw[black] (-5.167207792207792, 1.220779220779221) circle (2pt);
\filldraw[black] (1.7970779220779214, 0.38311688311688297) circle (2pt);
\filldraw[black] (2.27435064935065, 0.5194805194805197) circle (2pt);

\draw[|-|, black, draw opacity=1, line width=1] (-5.2,0) -- (4.85,0);
\node[below] at (-0.35,-0.1) {$w(I)$};

\end{tikzpicture}
}
    \vspace{-10mm}
    \caption{The line between the leftmost and rightmost terminals is fully covered by pyramids. Therefore, the width of the instance is bounded by the sum of the widths of all pyramids.}
    \label{fig:pyramids}
\end{figure}
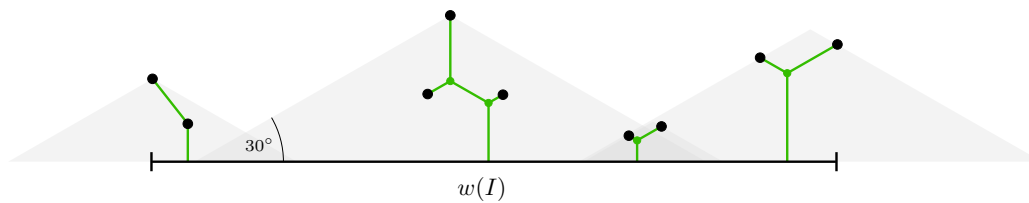

\begin{lemma}[Lower Bound]\label{lowerBound}
Let $I$ be an instance of the \esfl\ with set of terminals $R = \{(x_1, y_1), \dots, (x_n, y_n)\}$ sorted in non-decreasing order by $x_i$ and satisfying the property of Lemma \ref{lemmaDecomp}. Then, it holds that
\[
W_I=w(R) \leq \left(1 + \frac{2}{\tan(\ang{30})}\right) \cdot \mathrm{OPT}_I.
\]
\end{lemma}
\begin{proof}
    Let $T = (V \supseteq (R \cup \{\gamma\}), E)$ be an optimal solution for the instance $I$, and consider the connected components $C_1, \dots, C_k$ of the graph $T[V \setminus \{\gamma\}]$. For a component $C \in \{C_1, \dots, C_k\}$, let $r_C := \text{argmax}_{j \in V(C)} (x_j + \frac{y_j}{\tan(\ang{30})})$ and $l_C := \text{argmin}_{j \in V(C)} (x_j - \frac{y_j}{\tan(\ang{30})})$, where $V(C)$ is the set of vertices in the component.
We define 
\[P(C) := \{(x, y) \in \mathbb{R}^2 \mid y \geq 0, \ y \leq y_{l_C} - \tan(\ang{30})(x - x_{l_C}), \ y \leq y_{r_C} + \tan(\ang{30})(x - x_{r_C}) \}\] 
as the \emph{pyramid} of $C$ (see Figure \ref{fig:pyramid}). Here, $l_C$ lies on the left boundary, $r_C$ on the right boundary, and all other points in $C$ lie on the boundary or within the pyramid.

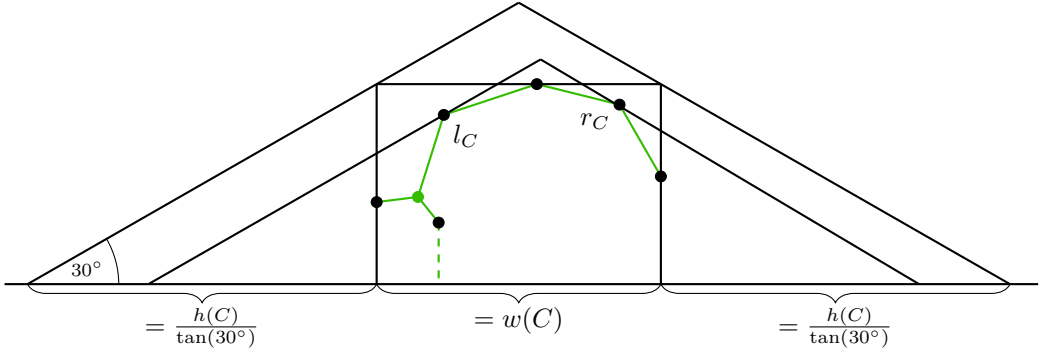
\begin{figure}[htb]
    \centering
    \begin{tikzpicture}[scale=0.68]
\definecolor{dgreen}{RGB}{52, 189, 0} 

    \coordinate (A) at (8.455,0);
    \coordinate (B) at (-10.555,0);
    \coordinate (C) at (-1.05,5.4877);
    \coordinate (D) at (-3.8, 1.6);
    \coordinate (E) at (-2.5, 3.3);
    \coordinate (F) at (-0.7, 3.9);
    \coordinate (G) at (0.9, 3.5);
    \coordinate (H) at (1.7, 2.1);
    \coordinate (I) at (-8.21577, 0);
    \coordinate (J) at (-0.626795, 4.3815);
    \coordinate (K) at (6.6962, 0);
    \coordinate (L) at (-2.6, 1.2);
    \coordinate (S) at (-3, 1.7);

    \pic[draw, angle eccentricity=1.2, angle radius=1.2cm] {angle=K--B--C};
    \node[] at (-9.4,0.3) {$\scriptstyle\ang{30}$};
        
    \draw[dgreen, dashed, thick] (L) -- (-2.6, 0);
    \draw[dgreen, thick] (L) -- (S);
    \draw[dgreen, thick] (D) -- (S);
    \draw[dgreen, thick] (E) -- (S);
    \draw[dgreen, thick] (E) -- (F);
    \draw[dgreen, thick] (G) -- (F);
    \draw[dgreen, thick] (G) -- (H);

    \draw[thick] (A) -- (C);
    \draw[thick] (-11, 0) -- (9, 0);

    \draw[thick] (-3.8, 0) -- (-3.8, 3.9);
    \draw[thick] (1.7, 3.9) -- (-3.8, 3.9);
    \draw[thick] (1.7, 0) -- (1.7, 3.9);
    
    \draw [decorate,decoration={brace,mirror,amplitude=7pt}] (-3.8,0) -- (1.7,0) 
        node[midway, below=5pt] {$=w(C)$};

    \draw [decorate,decoration={brace,mirror,amplitude=7pt}] (1.7,0) -- (A) 
        node[midway, below=5pt] {$= \frac{h(C)}{\tan(\ang{30})}$};

    \draw [decorate,decoration={brace,mirror,amplitude=7pt}] (B) -- (-3.8,0) 
        node[midway, below=5pt] {$= \frac{h(C)}{\tan(\ang{30})}$};
    
    \draw[thick] (C) -- (B);
    \draw[thick] (I) -- (J);
    \draw[thick] (J) -- (K);

    \filldraw (D) circle (3pt);
    \filldraw (E) circle (3pt);
    \filldraw (F) circle (3pt);
    \filldraw (G) circle (3pt);
    \filldraw (H) circle (3pt);
    \filldraw (L) circle (3pt);
    \filldraw[dgreen] (S) circle (3pt);

    \node[below right] at (E) {$l_C$};
    \node[below left] at (G) {$r_C$};
    
\end{tikzpicture}
    \caption{Depiction of the inner pyramid $P(C)$ and outer pyramid $P_{\scriptscriptstyle\square}(C)$ for a connected component $C$ from Lemma \ref{lowerBound}. The edge that connects the component to $\gamma$ is shown dashed. The width of $P(C)$ is bounded by $w(C) + \frac{2h(C)}{\tan(\ang{30})}$.}
    \label{fig:pyramid}
\end{figure}

We first show that the line in between $x_1$ and $x_n$ is fully covered by the pyramids of all components (see Figure \ref{fig:pyramids}). Consider two arbitrary consecutive terminals $x_i$ and $x_{i+1}$. By Lemma \ref{lemmaDecomp}, it holds that $a := \max_{j \in \{1, \dots, i\}} (x_j + \frac{y_{j}}{\tan(\ang{30})}) \geq \min_{j \in \{i+1, \dots, n\}} (x_j - \frac{y_{j}}{\tan(\ang{30})}) =: b$. The terminals $j_a$ and $j_b$, which attain the respective maximum and minimum, each belong to a component. Since $x_{j_a} \leq x_i$ and $x_{j_b} \geq x_{i+1}$, the left boundary of the pyramid for the component of $j_a$ lies to the left of $x_i$, and the right boundary of the pyramid for the component of $j_b$ lies to the right of $x_{i+1}$. If $j_a$ and $j_b$ are in the same component, then its pyramid trivially covers the segment between $x_i$ and $x_{i+1}$. If $j_a$ and $j_b$ are in different components, their pyramids must intersect due to $a \geq b$, thus fully covering the segment between $x_i$ and $x_{i+1}$.

Next, we establish an upper bound for the width of the pyramid $P(C)$ for each component $C$.
Consider the axis-aligned rectangle with size $w(C) \times h(C)$ that encloses $C$ and touches the line, i.e., a downwards-extended bounding box of $C$. We observe that the pyramid $P_{\scriptscriptstyle\square}(C)$, defined by the upper corners of this rectangle, fully contains $P(C)$ and therefore has a width at least as large as that of $P(C)$. 
Since the horizontal distance of the rectangle to the boundary of $P_{\scriptscriptstyle\square}(C)$ is $\frac{h(C)}{\tan(\ang{30})}$ on each side, the total width of $P_{\scriptscriptstyle\square}(C)$ is $w(C) + \frac{2h(C)}{\tan(\ang{30})}$. This therefore serves as an upper bound for the width of $P(C)$.

Since the line between $x_1$ and $x_n$ is fully covered by pyramids, it follows that the width $w(R)$ of the instance is bounded by the sum of the widths of all pyramids. Given that $\sum_{i=1}^k w(C_i) \leq \mathrm{OPT}_I$ and $\sum_{i=1}^k h(C_i) \leq \mathrm{OPT}_I$, we conclude that

\[
w(R) \leq \sum_{i=1}^k w(P(C_i)) \leq \sum_{i=1}^k \left( w(C_i) + \frac{2h(C_i)}{\tan(\ang{30})} \right) \leq \left(1 + \frac{2}{\tan(\ang{30})}\right) \cdot \mathrm{OPT}_I.
\]
\end{proof}

\subsection{Algorithm}
For transforming an \esfl-instance into an \est-instance, we use the straightforward approach of replacing the line with a sufficient number of terminal points.

\begin{definition}
For an \esfl-instance $I=(R=\{(x_1,y_1),\dots,(x_n,y_n)\},\gamma)$ and $\varepsilon>0$, let $I_{\varepsilon} := (R \cup L )$ be an instance for the \est, where $L$ is a set of $\ceil{\frac{n}{\varepsilon}}+1$ points, equally spaced along $\gamma$ between the leftmost and rightmost terminals. Thus, the distance between neighboring points is given by $d := W_I/\ceil{\frac{n}{\varepsilon}}$. More precisely, $L = \{l_i \mid i \in \{0, \dots, \ceil{\frac{n}{\varepsilon}}\}$, where $l_i := (x_{min}+i \cdot d, 0)$ and $x_{min} := \min_{(x, y) \in R}x$. We refer to $L$ as the set of \textbf{line points}, $R$ as the set of \textbf{real terminals}, and $S$ as the set of \textbf{Steiner points} in a given solution. Finally, we call edges between line points \textbf{line segments}, and the absence of a \emph{line segment} is referred to as a \textbf{hole}.
\end{definition}
See Figures \ref{fig:defI} and \ref{fig:defIEps} for reference. Real terminals are shown in black, Steiner points in green, and line points are depicted with a black border and white interior throughout the paper.

\begin{figure}[htb]
    \centering
    \begin{minipage}{.5\textwidth}
      \centering
      \begin{tikzpicture}[scale=2.6]

\draw [color={rgb, 255:red, 52; green, 189; blue, 0 }  ,draw opacity=1, line width=1]   (0.23, 0.00) -- (0.22, 0.01);
\draw [color={rgb, 255:red, 52; green, 189; blue, 0 }  ,draw opacity=1, line width=1]   (0.22, 0) -- (0.22, 0.78);
\draw [color={rgb, 255:red, 52; green, 189; blue, 0 }  ,draw opacity=1, line width=1]   (0.22, 0.78) -- (-0.50, 1.20);
\draw [color={rgb, 255:red, 52; green, 189; blue, 0 }  ,draw opacity=1, line width=1]   (0.22, 0.78) -- (0.60, 1.00);
\filldraw[black] (-0.50, 1.20) circle (1pt);
\filldraw[black] (0.60, 1.00) circle (1pt);
\filldraw[color={rgb, 255:red, 52; green, 189; blue, 0 }] (0.22, 0.78) circle (1pt);

\node at (0.05,-0.2) {Line};

\draw[black, draw opacity=1, line width=1] (-1,0) -- (1,0) ;

\end{tikzpicture}
      \caption{$I$ with optimal solution}
      \label{fig:defI}
    \end{minipage}%
    \begin{minipage}{.5\textwidth}
      \centering
      \begin{tikzpicture}[scale=2.6]                           

\draw[black, draw opacity=1, line width=1] (-1,0) -- (-0.5,0) ;
\draw[black, draw opacity=1, line width=1] (0.6,0) -- (1,0) ;
\draw [color={rgb, 255:red, 52; green, 189; blue, 0 }  ,draw opacity=1, line width=1]   (0.23, 0.00) -- (0.08, 0.10);
\draw [color={rgb, 255:red, 52; green, 189; blue, 0 }  ,draw opacity=1, line width=1]   (0.08, 0.10) -- (-0.13, 0.00);
\draw [color={rgb, 255:red, 52; green, 189; blue, 0 }  ,draw opacity=1, line width=1]   (0.08, 0.10) -- (0.13, 0.77);
\draw [color={rgb, 255:red, 52; green, 189; blue, 0 }  ,draw opacity=1, line width=1]   (0.13, 0.77) -- (-0.50, 1.20);
\draw [color={rgb, 255:red, 52; green, 189; blue, 0 }  ,draw opacity=1, line width=1]   (0.13, 0.77) -- (0.60, 1.00);
\draw [color={rgb, 255:red, 52; green, 189; blue, 0 }  ,draw opacity=1, line width=1]   (-0.50, 0.00) -- (-0.13, 0.00);
\draw [color={rgb, 255:red, 52; green, 189; blue, 0 }  ,draw opacity=1, line width=1]   (0.23, 0.00) -- (0.60, 0.00);
\filldraw[black] (-0.50, 1.20) circle (1pt);
\filldraw[black] (0.60, 1.00) circle (1pt);
\filldraw[black, fill=white, line width=0.4mm] (-0.50, 0.00) circle (1pt);
\filldraw[black, fill=white, line width=0.4mm] (-0.13, 0.00) circle (1pt);
\filldraw[black, fill=white, line width=0.4mm] (0.23, 0.00) circle (1pt);
\filldraw[black, fill=white, line width=0.4mm] (0.60, 0.00) circle (1pt);
\filldraw[color={rgb, 255:red, 52; green, 189; blue, 0 }] (0.08, 0.10) circle (1pt);
\filldraw[color={rgb, 255:red, 52; green, 189; blue, 0 }] (0.13, 0.77) circle (1pt);

\node at (0.07,-0.2) {hole};
\node at (-0.5,0.2) {line segment};
\node at (0.7,0.2) {line point};
\node at (1.1,1.1) {real terminal};

\end{tikzpicture}
      \caption{$I_\varepsilon$ with optimal solution}
      \label{fig:defIEps}
    \end{minipage}
\end{figure}

\begin{remark}
Unfortunately, the na\"\i ve approach of simply replacing the line with many terminal points and then using a PTAS for the \est\ is insufficient, as the resulting solution to the \esfl\ can still be very poor. As illustrated in Figure \ref{fig:poorapproxsolution}, the \est\ approximation lacks any incentive to use edges that run on the line $\gamma$, i.e., the line segments. Note that since we are using a black-box \est-PTAS, the approximate solution does not have to meet any specific structural properties, particularly not the angle condition. Even with arbitrarily tight approximations guarantees for the \est, the algorithm may output a solution that completely disregards the line. This is problematic because line segments are cost-free in the \esfl-instance, so the solution should contain as many of them as possible.

Luckily, using some post-processing on the solution, we can reduce the number of holes and bound it linearly in the number of real terminals in Lemma \ref{lemma:maxholes}. This routine, called \textsc{FillHoles} (Algorithm \ref{alg:transform}), employs two main strategies to fill holes without increasing the \est-solution weight. The first strategy involves inserting a previouly missing line segment and removing the longest edge of the resulting cycle (see Figure \ref{fig:fillHolesStep3}). The second strategy computes local optima on specific substructures (see Figure \ref{fig:fillHolesStep4}). Each time a hole is filled, the algorithm restarts from the beginning to ensure a particular solution structure.
Although this makes it not immediately clear that the algorithm terminates, we will use these properties to prove its termination and analyze its runtime in Lemma \ref{lemma:runtime}, leveraging structural insights from Lemma \ref{lemma:4lines}.

We arrive at Algorithm \ref{alg:ptas}, which begins by constructing the \est-instance $I_\varepsilon$ by incorporating the line points $L$. It then uses a \est-PTAS $A$ to obtain an approximative solution $T_\varepsilon$ to this instance. This solution is then further refined by \textsc{FillHoles} to ensure it contains a sufficient number of line segments (as outlined above). Next, all line points are then contracted to $\gamma$, and any cycles introduced during this process are removed. Finally, the resulting solution is returned.
\end{remark}

\begin{algorithm}[p]
\normalsize
\caption{\esfl\ approximation scheme}
\textbf{Input:} \emph{\esfl-Instance $I=(R \subset \mathbb{R}^2, \gamma=\{(x,0)\mid x\in \mathbb{R}\})$ and $\varepsilon > 0$; \est-PTAS A}
\vspace{-2mm}
\begin{pseudo}
Construct $I_\varepsilon = (R \cup L)$ and obtain $(1+\varepsilon)$-approximation $T_\varepsilon = (V = R \cup L \cup S, E)$ \\*&
using algorithm $A$. \\
$T_\varepsilon := $ \textsc{FillHoles($T_\varepsilon$)}\\
Constract all line points $L$ in $T_{\varepsilon}$ to $\gamma$. \\
Delete any cycles and return the resulting \esfl-solution: \\*&
\kw{return} $T= \textsc{Mst($T_\varepsilon$)}$
\end{pseudo}
\vspace{-3mm}
\label{alg:ptas}
\end{algorithm}

\begin{algorithm}[p]
\caption{\textsc{FillHoles}}
\textbf{Input:} $T_\varepsilon$ where $V(T_\varepsilon)$ are the nodes, $S(T_\varepsilon)\subset V(T_\varepsilon)$ the Steiner points, $L(T_\varepsilon)$ the line points, and $E(T_\varepsilon)$ the edges of $T_\varepsilon$.
\vspace{-2mm}
\begin{pseudo}
Delete all Steiner points with degree 2 and connect their neighbors directly.\\
Split Steiner points with degree greater than $3$ into several points, each with degree $3$:\\*&
\kw{while} $\exists v \in S(T_\varepsilon) \text{ with } |\delta(v)| > 3$ \kw{do} \\+*&
Let $\{v, w_1\}, \{v, w_2\} \in \delta(v) \text{ and } v = (x_v, y_v)$ \\*&
Let $u = (x_v, y_v)$ be a new Steiner point \\*&
$S(T_\varepsilon) := S(T_\varepsilon) \cup \{u\}$\\*&
$E(T_\varepsilon) := (E(T_\varepsilon) \cup \{\{u, v\}, \{u, w_1\}, \{u, w_2\}\}) \setminus \{\{v, w_1\}, \{v, w_2\}\}$\\--
Try to fill holes by swapping a missing line segment with a non line-segment that has\\*&larger or equal length (see Figure \ref{fig:fillHolesStep3}):\\*&
\kw{for} $i := 0$ \kw{to} $\ceil{\frac{n}{\varepsilon}}-1$ \kw{do} \\+*&
\kw{if} $\{l_i, l_{i+1}\} \notin E(T_\varepsilon)$ \kw{then} \\+*&
Choose the largest edge $e_{max}$ of the cycle in the graph\\*&$G'=(V(T_\varepsilon), E(T_\varepsilon) \cup \{\{l_i, l_{i+1}\}\})$
(in case of equality, let $e_{max}$ be an edge\\*&that is not a line segment)\\*&
\kw{if} $e_{max} \textit{ is not a line segment } $\kw{then}\\+*&
$E(T_\varepsilon) := (E(T_\varepsilon) \cup \{\{l_i,l_{i+1}\}\}) \setminus \{e_{max}\}$\\*&
\kw{goto} Step 1 {\color{darkgray} // A hole was filled} \\---
Remove paths of Steiner points of length $5$, where each Steiner point shares an edge\\*&with a line point, by computing a local optimum (see Figure \ref{fig:fillHolesStep4}):\\*&
$\kw{if } \exists \text{ Path } P=(s_1,\dots,s_5) \text{ in } T_\varepsilon \text{ with } s_i\in S(T_\varepsilon) \text{ and distinct line points }$\\*&$l'_1,\dots,l'_5 \in L(T_\varepsilon)$ $\text{ such that } \{l'_i,s_i\}\in E(T_\varepsilon) \text{ for } i\in \{1,\dots,5\}$ \textbf{then}\\+*&
Consider the \est-subinstance $J = (l'_1,\dots,l'_5, s_1, s_5)$ \\*& 
Compute an optimal solution $T^*$ to $J$ by using Melzak's algorithm~\cite{melzak1961}. \\*&
$T_\varepsilon := T_\varepsilon[V(T_\varepsilon) \setminus \{s_2, \dots, s_4\}]$\\*&
$S(T_\varepsilon) := S(T_\varepsilon) \cup S(T^*)$\\*&
$E(T_\varepsilon) := E(T_\varepsilon) \cup E(T^*)$\\*&
\kw{goto} Step 1 {\color{darkgray} // A hole was filled}\\-
\kw{return} $T_\varepsilon$
\end{pseudo}
\vspace{-3mm}
\label{alg:transform}
\end{algorithm}
\begin{figure}[h!]
    \centering
    \begin{subfigure}[t]{0.3\textwidth}
        \centering
        \begin{tikzpicture}[scale=0.75]
\definecolor{dgreen}{RGB}{52, 189, 0} 
\draw[color=white, thick, rounded corners=4pt] 
        (0.7, -0.2) -- (5.3, -0.2) -- (5.1, 0.7) -- (4.8, 0.7) -- (4.4, 0.2) -- (1.6, 0.2) -- (1.2, 0.7) -- (0.9, 0.7) -- cycle;
\draw [color=dgreen  ,draw opacity=1, line width=1]   (1.00, 0.00) -- (1.10, 0.40);
\draw [color=dgreen  ,draw opacity=1, line width=1]   (2.00, 0.00) -- (2.05, 0.40);
\draw [color=dgreen  ,draw opacity=1, line width=1]   (3.00, 0.00) -- (3.00, 0.40);
\draw [color=dgreen  ,draw opacity=1, line width=1]   (4.00, 0.00) -- (3.95, 0.40);
\draw [color=dgreen  ,draw opacity=1, line width=1]   (5.00, 0.00) -- (4.90, 0.40);
\draw [color=dgreen  ,draw opacity=1, line width=1]   (1.10, 0.40) -- (2.05, 0.40);
\draw [color=dgreen  ,draw opacity=1, line width=1]   (2.05, 0.40) -- (3.00, 0.40);
\draw [color=dgreen  ,draw opacity=1, line width=1]   (3.00, 0.40) -- (3.95, 0.40);
\draw [color=dgreen  ,draw opacity=1, line width=1]   (3.95, 0.40) -- (4.90, 0.40);
\draw [color=dgreen  ,draw opacity=1, line width=1,dashed]   (0.70, 1.00) -- (1.10, 0.40);
\draw [color=dgreen  ,draw opacity=1, line width=1,dashed]   (4.90, 0.40) -- (5.30, 1.00);
\filldraw[color=black, fill=white, line width=0.4mm] (1.00, 0.00) circle (2.5pt);
\filldraw[color=black, fill=white, line width=0.4mm] (2.00, 0.00) circle (2.5pt);
\filldraw[color=black, fill=white, line width=0.4mm] (3.00, 0.00) circle (2.5pt);
\filldraw[color=black, fill=white, line width=0.4mm] (4.00, 0.00) circle (2.5pt);
\filldraw[color=black, fill=white, line width=0.4mm] (5.00, 0.00) circle (2.5pt);

\filldraw[color=dgreen] (1.10, 0.40) circle (2.5pt);
\filldraw[color=dgreen] (2.05, 0.40) circle (2.5pt);
\filldraw[color=dgreen] (3.00, 0.40) circle (2.5pt);
\filldraw[color=dgreen] (3.95, 0.40) circle (2.5pt);
\filldraw[color=dgreen] (4.90, 0.40) circle (2.5pt);
\end{tikzpicture}
        \caption{A possible approximation of ${I_\varepsilon}$ that needs to be post-processed.}
        \label{fig:poorapproxsolution}
    \end{subfigure}
    \hfill
    \begin{subfigure}[t]{0.3\textwidth}
        \centering
                \begin{tikzpicture}[scale=0.75]
\definecolor{dgreen}{RGB}{52, 189, 0} 
\draw [color=dgreen  ,draw opacity=1, line width=1]   (1.00, 0.00) -- (1.10, 0.40);
\draw [color=dgreen  ,draw opacity=1, line width=1]   (2.00, 0.00) -- (2.05, 0.40);
\draw [color=dgreen  ,draw opacity=1, line width=1]   (3.00, 0.00) -- (3.00, 0.40);
\draw [color=dgreen  ,draw opacity=1, line width=1]   (4.00, 0.00) -- (3.95, 0.40);
\draw [color=dgreen  ,draw opacity=1, line width=1]   (5.00, 0.00) -- (4.90, 0.40);
\draw [color=dgreen  ,draw opacity=1, line width=1]   (1.10, 0.40) -- (2.05, 0.40);
\draw [color=dgreen  ,draw opacity=1, line width=1]   (2.05, 0.40) -- (3.00, 0.40);
\draw [color=dgreen  ,draw opacity=1, line width=1]   (3.00, 0.40) -- (3.95, 0.40);
\draw [color=dgreen  ,draw opacity=1, line width=1]   (3.95, 0.40) -- (4.90, 0.40);
\draw [color=dgreen  ,draw opacity=1, line width=1,dashed]   (0.70, 1.00) -- (1.10, 0.40);
\draw [color=dgreen  ,draw opacity=1, line width=1,dashed]   (4.90, 0.40) -- (5.30, 1.00);
\filldraw[color=black, fill=white, line width=0.4mm] (1.00, 0.00) circle (2.5pt);
\filldraw[color=black, fill=white, line width=0.4mm] (2.00, 0.00) circle (2.5pt);
\filldraw[color=black, fill=white, line width=0.4mm] (3.00, 0.00) circle (2.5pt);
\filldraw[color=black, fill=white, line width=0.4mm] (4.00, 0.00) circle (2.5pt);
\filldraw[color=black, fill=white, line width=0.4mm] (5.00, 0.00) circle (2.5pt);

\filldraw[color=dgreen] (1.10, 0.40) circle (2.5pt);
\filldraw[color=dgreen] (2.05, 0.40) circle (2.5pt);
\filldraw[color=dgreen] (3.00, 0.40) circle (2.5pt);
\filldraw[color=dgreen] (3.95, 0.40) circle (2.5pt);
\filldraw[color=dgreen] (4.90, 0.40) circle (2.5pt);
\draw[color=blue, thick, rounded corners=4pt] 
        (0.7, -0.2) -- (5.3, -0.2) -- (5.1, 0.7) -- (4.8, 0.7) -- (4.4, 0.2) -- (1.6, 0.2) -- (1.2, 0.7) -- (0.9, 0.7) -- cycle;

\end{tikzpicture}
        \caption{In Step~4, a path of $5$ Steiner points, where each point has an edge to a line point, is found, and a subinstance $J$ is constructed (highlighted in blue).}
    \end{subfigure}
    \hfill
    \begin{subfigure}[t]{0.3\textwidth}
        \centering
                \begin{tikzpicture}[scale=0.75]
\definecolor{dgreen}{RGB}{52, 189, 0} 
\draw [color={rgb, 255:red, 52; green, 189; blue, 0 }  ,draw opacity=1, line width=1]   (1.17, 0.18) -- (1.00, 0.00);
\draw [color={rgb, 255:red, 52; green, 189; blue, 0 }  ,draw opacity=1, line width=1]   (1.17, 0.18) -- (1.10, 0.40);
\draw [color={rgb, 255:red, 52; green, 189; blue, 0 }  ,draw opacity=1, line width=1]   (5.00, 0.00) -- (4.83, 0.18);
\draw [color={rgb, 255:red, 52; green, 189; blue, 0 }  ,draw opacity=1, line width=1]   (4.83, 0.18) -- (4.00, 0.00);
\draw [color={rgb, 255:red, 52; green, 189; blue, 0 }  ,draw opacity=1, line width=1]   (4.83, 0.18) -- (4.90, 0.40);
\draw [color={rgb, 255:red, 52; green, 189; blue, 0 }  ,draw opacity=1, line width=1]   (2.00, 0.00) -- (3.00, 0.00);
\draw [color={rgb, 255:red, 52; green, 189; blue, 0 }  ,draw opacity=1, line width=1]   (3.00, 0.00) -- (4.00, 0.00);
\draw [color={rgb, 255:red, 52; green, 189; blue, 0 }  ,draw opacity=1, line width=1]   (1.17, 0.18) -- (2,0);

\draw [color=dgreen  ,draw opacity=1, line width=1,dashed]   (0.70, 1.00) -- (1.10, 0.40);
\draw [color=dgreen  ,draw opacity=1, line width=1,dashed]   (4.90, 0.40) -- (5.30, 1.00);

\filldraw[color=dgreen] (1.10, 0.40) circle (2.5pt);
\filldraw[color=dgreen] (4.90, 0.40) circle (2.5pt);
\filldraw[color=black, fill=white, line width=0.4mm] (1.00, 0.00) circle (2.5pt);
\filldraw[color=black, fill=white, line width=0.4mm] (2.00, 0.00) circle (2.5pt);
\filldraw[color=black, fill=white, line width=0.4mm] (3.00, 0.00) circle (2.5pt);
\filldraw[color=black, fill=white, line width=0.4mm] (4.00, 0.00) circle (2.5pt);
\filldraw[color=black, fill=white, line width=0.4mm] (5.00, 0.00) circle (2.5pt);
\filldraw[color={rgb, 255:red, 52; green, 189; blue, 0 }] (1.17, 0.18) circle (2.5pt);
\filldraw[color={rgb, 255:red, 52; green, 189; blue, 0 }] (4.83, 0.18) circle (2.5pt);

\draw[color=white, thick, rounded corners=4pt] 
        (0.7, -0.2) -- (5.3, -0.2) -- (5.1, 0.7) -- (4.8, 0.7) -- (4.4, 0.2) -- (1.6, 0.2) -- (1.2, 0.7) -- (0.9, 0.7) -- cycle;

\end{tikzpicture}
        \caption{The redundant Steiner points are replaced by the local optimum of subinstance $J$. The algorithm jumps back to Step~1 to remove the two Steiner points with degree 2.}
    \end{subfigure}
    \caption{An illustration of \textsc{FillHoles} (Algorithm \ref{alg:transform}), post-processing a potential approximate solution for $I_\varepsilon$ to reduce the number of holes.}
    \label{fig:fillHolesStep4}
\end{figure}
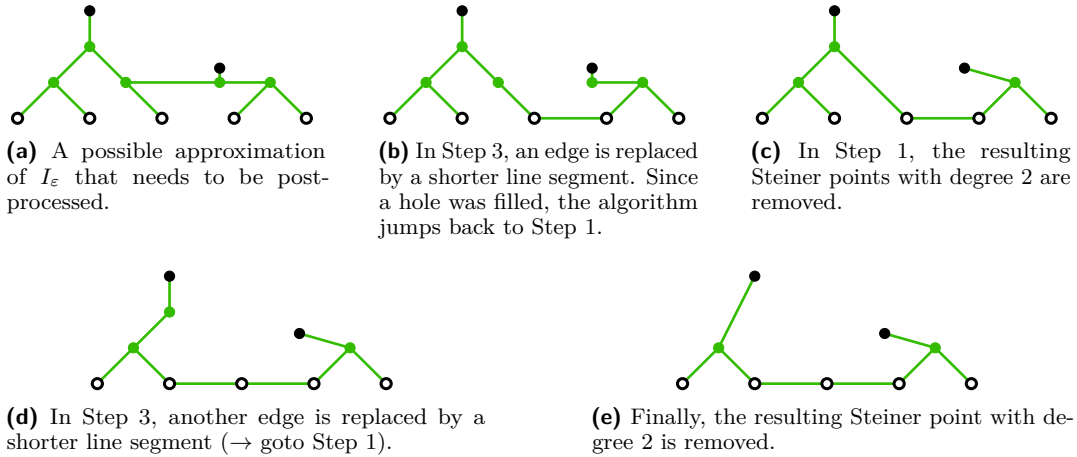
\begin{figure}[h!]
    \centering
    \begin{subfigure}[t]{0.3\textwidth}
        \centering
        \begin{tikzpicture}[scale=0.95]
\definecolor{dgreen}{RGB}{52, 189, 0} 
\draw [color=dgreen  ,draw opacity=1, line width=1]   (1.00, 1.50) -- (1.00, 1.00);
\draw [color=dgreen  ,draw opacity=1, line width=1]   (2.80, 0.70) -- (2.80, 0.50);
\draw [color=dgreen  ,draw opacity=1, line width=1]   (0.00, 0.00) -- (0.50, 0.50);
\draw [color=dgreen  ,draw opacity=1, line width=1]   (1.00, 0.00) -- (0.50, 0.50);
\draw [color=dgreen  ,draw opacity=1, line width=1]   (2.00, 0.00) -- (1.50, 0.50);
\draw [color=dgreen  ,draw opacity=1, line width=1]   (3.00, 0.00) -- (3.50, 0.50);
\draw [color=dgreen  ,draw opacity=1, line width=1]   (4.00, 0.00) -- (3.50, 0.50);
\draw [color=dgreen  ,draw opacity=1, line width=1]   (0.50, 0.50) -- (1.00, 1.00);
\draw [color=dgreen  ,draw opacity=1, line width=1]   (1.50, 0.50) -- (1.00, 1.00);
\draw [color=dgreen  ,draw opacity=1, line width=1]   (1.50, 0.50) -- (2.80, 0.50);
\draw [color=dgreen  ,draw opacity=1, line width=1]   (2.80, 0.50) -- (3.50, 0.50);
\filldraw[black] (1.00, 1.50) circle (2pt);
\filldraw[black] (2.80, 0.70) circle (2pt);
\filldraw[color=black, fill=white, line width=0.4mm] (0.00, 0.00) circle (2pt);
\filldraw[color=black, fill=white, line width=0.4mm] (1.00, 0.00) circle (2pt);
\filldraw[color=black, fill=white, line width=0.4mm] (2.00, 0.00) circle (2pt);
\filldraw[color=black, fill=white, line width=0.4mm] (3.00, 0.00) circle (2pt);
\filldraw[color=black, fill=white, line width=0.4mm] (4.00, 0.00) circle (2pt);
\filldraw[color=dgreen] (0.50, 0.50) circle (2pt);
\filldraw[color=dgreen] (1.50, 0.50) circle (2pt);
\filldraw[color=dgreen] (2.80, 0.50) circle (2pt);
\filldraw[color=dgreen] (3.50, 0.50) circle (2pt);
\filldraw[color=dgreen] (1.00, 1.00) circle (2pt);

\end{tikzpicture}
        \caption{A possible approximation of ${I_\varepsilon}$ that needs to be post-processed.}
    \end{subfigure}
    \hfill
    \begin{subfigure}[t]{0.3\textwidth}
        \centering
                \begin{tikzpicture}[scale=0.95]
\definecolor{dgreen}{RGB}{52, 189, 0} 
\draw [color=dgreen  ,draw opacity=1, line width=1]   (1.00, 1.50) -- (1.00, 1.00);
\draw [color=dgreen  ,draw opacity=1, line width=1]   (2.80, 0.70) -- (2.80, 0.50);
\draw [color=dgreen  ,draw opacity=1, line width=1]   (0.00, 0.00) -- (0.50, 0.50);
\draw [color=dgreen  ,draw opacity=1, line width=1]   (1.00, 0.00) -- (0.50, 0.50);
\draw [color=dgreen  ,draw opacity=1, line width=1]   (2.00, 0.00) -- (1.50, 0.50);
\draw [color=dgreen  ,draw opacity=1, line width=1]   (3.00, 0.00) -- (3.50, 0.50);
\draw [color=dgreen  ,draw opacity=1, line width=1]   (4.00, 0.00) -- (3.50, 0.50);
\draw [color=dgreen  ,draw opacity=1, line width=1]   (0.50, 0.50) -- (1.00, 1.00);
\draw [color=dgreen  ,draw opacity=1, line width=1]   (1.50, 0.50) -- (1.00, 1.00);
\draw [color=dgreen  ,draw opacity=1, line width=1]   (2.0, 0.00) -- (3.0, 0.00);
\draw [color=dgreen  ,draw opacity=1, line width=1]   (2.80, 0.50) -- (3.50, 0.50);
\filldraw[black] (1.00, 1.50) circle (2pt);
\filldraw[black] (2.80, 0.70) circle (2pt);
\filldraw[color=black, fill=white, line width=0.4mm] (0.00, 0.00) circle (2pt);
\filldraw[color=black, fill=white, line width=0.4mm] (1.00, 0.00) circle (2pt);
\filldraw[color=black, fill=white, line width=0.4mm] (2.00, 0.00) circle (2pt);
\filldraw[color=black, fill=white, line width=0.4mm] (3.00, 0.00) circle (2pt);
\filldraw[color=black, fill=white, line width=0.4mm] (4.00, 0.00) circle (2pt);
\filldraw[color=dgreen] (0.50, 0.50) circle (2pt);
\filldraw[color=dgreen] (1.50, 0.50) circle (2pt);
\filldraw[color=dgreen] (2.80, 0.50) circle (2pt);
\filldraw[color=dgreen] (3.50, 0.50) circle (2pt);
\filldraw[color=dgreen] (1.00, 1.00) circle (2pt);

\end{tikzpicture}
        \caption{In Step~3, an edge is replaced by a shorter line segment. Since a hole was filled, the algorithm jumps back to Step~1.}
    \end{subfigure}
    \hfill
    \begin{subfigure}[t]{0.3\textwidth}
        \centering
        \begin{tikzpicture}[scale=0.95]
\definecolor{dgreen}{RGB}{52, 189, 0} 
\draw [color=dgreen  ,draw opacity=1, line width=1]   (1.00, 1.50) -- (1.00, 1.00);
\draw [color=dgreen  ,draw opacity=1, line width=1]   (0.00, 0.00) -- (0.50, 0.50);
\draw [color=dgreen  ,draw opacity=1, line width=1]   (1.00, 0.00) -- (0.50, 0.50);
\draw [color=dgreen  ,draw opacity=1, line width=1]   (2.00, 0.00) -- (1.50, 0.50);
\draw [color=dgreen  ,draw opacity=1, line width=1]   (3.00, 0.00) -- (3.50, 0.50);
\draw [color=dgreen  ,draw opacity=1, line width=1]   (4.00, 0.00) -- (3.50, 0.50);
\draw [color=dgreen  ,draw opacity=1, line width=1]   (0.50, 0.50) -- (1.00, 1.00);
\draw [color=dgreen  ,draw opacity=1, line width=1]   (1.50, 0.50) -- (1.00, 1.00);
\draw [color=dgreen  ,draw opacity=1, line width=1]   (2.0, 0.00) -- (3.0, 0.00);
\draw [color=dgreen  ,draw opacity=1, line width=1]   (2.80, 0.70) -- (3.50, 0.50);
\filldraw[black] (1.00, 1.50) circle (2pt);
\filldraw[black] (2.80, 0.70) circle (2pt);
\filldraw[color=black, fill=white, line width=0.4mm] (0.00, 0.00) circle (2pt);
\filldraw[color=black, fill=white, line width=0.4mm] (1.00, 0.00) circle (2pt);
\filldraw[color=black, fill=white, line width=0.4mm] (2.00, 0.00) circle (2pt);
\filldraw[color=black, fill=white, line width=0.4mm] (3.00, 0.00) circle (2pt);
\filldraw[color=black, fill=white, line width=0.4mm] (4.00, 0.00) circle (2pt);
\filldraw[color=dgreen] (0.50, 0.50) circle (2pt);
\filldraw[color=dgreen] (3.50, 0.50) circle (2pt);
\filldraw[color=dgreen] (1.00, 1.00) circle (2pt);

\end{tikzpicture}
        \caption{In Step~1, the resulting Steiner points with degree $2$ are removed.}
    \end{subfigure}
    
    \vspace{1em} 
    \begin{subfigure}[t]{0.45\textwidth}
        \centering
        \begin{tikzpicture}[scale=0.95]
\definecolor{dgreen}{RGB}{52, 189, 0} 
\draw [color=dgreen  ,draw opacity=1, line width=1]   (1.00, 1.50) -- (1.00, 1.00);
\draw [color=dgreen  ,draw opacity=1, line width=1]   (0.00, 0.00) -- (0.50, 0.50);
\draw [color=dgreen  ,draw opacity=1, line width=1]   (1.00, 0.00) -- (0.50, 0.50);
\draw [color=dgreen  ,draw opacity=1, line width=1]   (3.00, 0.00) -- (3.50, 0.50);
\draw [color=dgreen  ,draw opacity=1, line width=1]   (4.00, 0.00) -- (3.50, 0.50);
\draw [color=dgreen  ,draw opacity=1, line width=1]   (0.50, 0.50) -- (1.00, 1.00);
\draw [color=dgreen  ,draw opacity=1, line width=1]   (1.00, 0.00) -- (2.00, 0.00);
\draw [color=dgreen  ,draw opacity=1, line width=1]   (2.0, 0.00) -- (3.0, 0.00);
\draw [color=dgreen  ,draw opacity=1, line width=1]   (2.80, 0.70) -- (3.50, 0.50);
\filldraw[black] (1.00, 1.50) circle (2pt);
\filldraw[black] (2.80, 0.70) circle (2pt);
\filldraw[color=black, fill=white, line width=0.4mm] (0.00, 0.00) circle (2pt);
\filldraw[color=black, fill=white, line width=0.4mm] (1.00, 0.00) circle (2pt);
\filldraw[color=black, fill=white, line width=0.4mm] (2.00, 0.00) circle (2pt);
\filldraw[color=black, fill=white, line width=0.4mm] (3.00, 0.00) circle (2pt);
\filldraw[color=black, fill=white, line width=0.4mm] (4.00, 0.00) circle (2pt);
\filldraw[color=dgreen] (0.50, 0.50) circle (2pt);
\filldraw[color=dgreen] (3.50, 0.50) circle (2pt);
\filldraw[color=dgreen] (1.00, 1.00) circle (2pt);

\end{tikzpicture}
        \caption{In Step~3, another edge is replaced by a shorter line segment ($\rightarrow$ goto Step~1).}
    \end{subfigure}
    \hfill
    \begin{subfigure}[t]{0.45\textwidth}
        \centering
        \begin{tikzpicture}[scale=0.95]
\definecolor{dgreen}{RGB}{52, 189, 0} 
\draw [color=dgreen  ,draw opacity=1, line width=1]   (1.00, 1.50) -- (0.50, 0.50);
\draw [color=dgreen  ,draw opacity=1, line width=1]   (2.80, 0.70) -- (3.50, 0.50);
\draw [color=dgreen  ,draw opacity=1, line width=1]   (0.00, 0.00) -- (0.50, 0.50);
\draw [color=dgreen  ,draw opacity=1, line width=1]   (1.00, 0.00) -- (0.50, 0.50);
\draw [color=dgreen  ,draw opacity=1, line width=1]   (1.00, 0.00) -- (2.00, 0.00);
\draw [color=dgreen  ,draw opacity=1, line width=1]   (2.00, 0.00) -- (3.00, 0.00);
\draw [color=dgreen  ,draw opacity=1, line width=1]   (3.00, 0.00) -- (3.50, 0.50);
\draw [color=dgreen  ,draw opacity=1, line width=1]   (4.00, 0.00) -- (3.50, 0.50);
\filldraw[black] (1.00, 1.50) circle (2pt);
\filldraw[black] (2.80, 0.70) circle (2pt);
\filldraw[color=black, fill=white, line width=0.4mm] (0.00, 0.00) circle (2pt);
\filldraw[color=black, fill=white, line width=0.4mm] (1.00, 0.00) circle (2pt);
\filldraw[color=black, fill=white, line width=0.4mm] (2.00, 0.00) circle (2pt);
\filldraw[color=black, fill=white, line width=0.4mm] (3.00, 0.00) circle (2pt);
\filldraw[color=black, fill=white, line width=0.4mm] (4.00, 0.00) circle (2pt);
\filldraw[color=dgreen] (0.50, 0.50) circle (2pt);
\filldraw[color=dgreen] (3.50, 0.50) circle (2pt);
\end{tikzpicture}
        \caption{Finally, the resulting Steiner point with degree $2$ is removed.}
    \end{subfigure}

    \caption{An illustration of \textsc{FillHoles} (Algorithm \ref{alg:transform}), post-processing a potential approximate solution for $I_\varepsilon$ to reduce the number of holes.}
    \label{fig:fillHolesStep3}
\end{figure}

\begin{lemma}\label{lemma:runtime}
    \textsc{FillHoles} (Algorithm \ref{alg:transform}) preserves the tree property of $T_\varepsilon$ and does not increase its weight. Under the assumption that $T_\varepsilon$ initially does not contain any Steiner points with degree less than 3, the runtime of \textsc{FillHoles} is bounded by $O((\frac{n}{\varepsilon})^3)$. 
\end{lemma}
\begin{proof}
Since we are working in the Euclidean metric, Steps 1--3 clearly preserve the tree property and do not increase the weight of $T_\varepsilon$. After Step~2, we can assume that each Steiner point in $T_\varepsilon$ has degree 3. Note that Step~3 retains this property, as each time Step~3 modifies the tree, it jumps back to Step~1. Consequently, when the Steiner points $s_2,\dots,s_4$ are removed and replaced with an optimal solution to the subinstance $J = (l'_1,\dots,l'_5, s_1, s_5)$ in Step~4, $T_\varepsilon$ remains a tree and does not increase in weight. 

For analyzing the runtime of \textsc{FillHoles}, we first bound how often \mbox{\textit{\textbf{goto} Step~1}} is invoked. Observe that no step can introduce a new hole and before every invocation of \mbox{\textit{\textbf{goto} Step 1}}, at least one hole is closed: Step~1 and 2 can only delete Steiner points and their incident edges and therefore cannot introduce a new line segment. Step~3 modifies the tree only if $e_{max}$ is not a line segment. Since in this case the line segment $\{l_i,l_{i+1}\}$ is added, the number of holes decreases by one before jumping back to Step~1.
Step~4 only removes Steiner points and their incident edges, so it also cannot introduce any new holes. Additionally, since $s_1,\dots,s_5$ form a path and $T_\varepsilon$ is a tree, none of $l'_1,\dots,l'_5$ can share a line segment. As we will show in Lemma \ref{lemma:4lines}, the local optimum $T^*$ contains at most 4 edges connecting points above $y=0$ to points on $y=0$. Therefore, $T^*$ must include at least one line segment, ensuring that Step~4 always closes at least one hole before jumping back to Step~1. We conclude that no step introduces a new hole and before every invocation of \mbox{\textit{\textbf{goto} Step~1}}, at least one hole is closed. Since the initial number of holes is bounded by $|L|-1=\ceil{\frac{n}{\varepsilon}}$, each step is executed at most $\ceil{\frac{n}{\varepsilon}}+1$ times. 

We now establish an upper bound on the size of $T_\varepsilon$. As introduced in Section \ref{sec:preliminaries}, a solution with $n$ terminals, where all Steiner points have degree 3, can include at most $n-2$ Steiner points. We can safely assume that $T_\varepsilon$ initially does not contain any Steiner points with a degree less than 3, as such points could easily be removed as a final step of the \est-PTAS without increasing runtime and solution weight. Since Steiner points with degree greater than 3 can be split into Steiner points with degree 3, the initial number of Steiner points in $T_\varepsilon$ is bounded by the number of terminals. Every time $T_\varepsilon$ is modified, the algorithm revisits Steps 1 and 2 to ensure that this property is maintained. Consequently, the maximum number of Steiner points with degree less than 3 in $T_\varepsilon$ is two. Specifically, two points with degree 2 may be present right after Step~3 (the incident nodes of $e_{max}$) or after Step~4 ($s_1$ and/or $s_5$). Therefore, at any given time, the number of nodes and edges in $T_\varepsilon$ is bounded by $O(\frac{n}{\varepsilon})$.

It follows that when the graph is represented as an adjacency list, Step~1 and 2 can be easily implemented in time $O(\frac{n}{\varepsilon})$. Step~3 runs for at most $\left\lceil\frac{n}{\varepsilon}\right\rceil$ iterations, where each iteration can be implemented using a Depth First Search. Therefore, the runtime of Step~3 is bounded by $O((\frac{n}{\varepsilon})^2)$. The challenging aspect of Step~4 is finding the specific substructure in $T_\varepsilon$ or concluding that none exists. For this, we consider the subgraph $T'$ of $T_\varepsilon$, induced by all Steiner points that share an edge with a line point. In $T'$, we find all connected components using Depth First Search. Since each node in $T_\varepsilon$ has degree 3 and each node in $T'$ has at least one edge to a line point, every node in $T'$ has at most degree 2. Thus, each connected component of $T'$ forms a path, allowing us to find suitable paths of size 5 in linear time. Once a suitable path is found, computing the local optimum requires only constant time using the algorithm of Melzak~\cite{melzak1961}, as the size of the subinstance $J$ is constant. Given that the runtime of each step is bounded by $O((\frac{n}{\varepsilon})^2)$ and each step is executed at most $\ceil{\frac{n}{\varepsilon}}+1$ times, the overall runtime of \textsc{FillHoles} is $O((\frac{n}{\varepsilon})^3)$.
\end{proof}

In the proof of Lemma \ref{lemma:runtime}, we relied on the fact that computing the local optimum in Step~4 always introduces at least one new line segment, i.e., closes a hole. To show this, we now establish a structural property of \est-solutions for instances that only contain line points and two real terminals. Specifically, we demonstrate that any optimal solution for such instances includes at most four edges connecting line points to non-line points.

\begin{restatable}{lemma}
{lemmaLineConnections}\label{lemma:4lines} 
    Let $T$ be an optimal solution to an instance of the \est\ that consists of two terminals with $y>0$ and all other terminals at $y=0$. Then, $T$ has at most $4$ edges connecting points with $y>0$ to points at $y=0$.
\end{restatable}
\begin{proof}
    Consider an instance with the terminal set $R=\{q,r\}\cup P$ with $q_y>0$, $r_y>0$ and $v_y=0$ for all $v\in P$. Let $T$ be an optimal solution to $I$. In the following, we assume that all edges are oriented downward, while horizontal edges remain undirected. Note that $T$ satisfies the degree and angle conditions for Steiner points, meaning each Steiner point is incident to exactly three edges that are equally spaced at $\ang{120}$-angles. We introduce the notations of \emph{strongly} and \emph{weakly} downward-directed edges: Edges whose smallest angle to the horizontal is greater than \ang{60} are called \emph{strongly} downward-directed. Edges whose smallest angle to the horizontal is less than or equal to \ang{60} and greater than \ang{0} are called \emph{weakly} downward-directed. Edges with an angle of \ang{0} are called horizontal. From the degree property of Steiner points, it follows that if a strongly downward-directed edge hits a Steiner point from above, this Steiner point is the source of two weakly downward-directed edges (\ref{fig:4lines:1}). Conversely, when a weakly downward-directed edge hits a Steiner point from above, it either combines with another weakly downward-directed edge to give rise to a strongly downward-directed edge (\ref{fig:4lines:2}) or splits into one weakly downward-directed edge and one horizontal edge (\ref{fig:4lines:3}). Note that each edge in $T$ classifies as either strongly or weakly downward-directed or horizontal (see Figure \ref{fig:4lines}).

    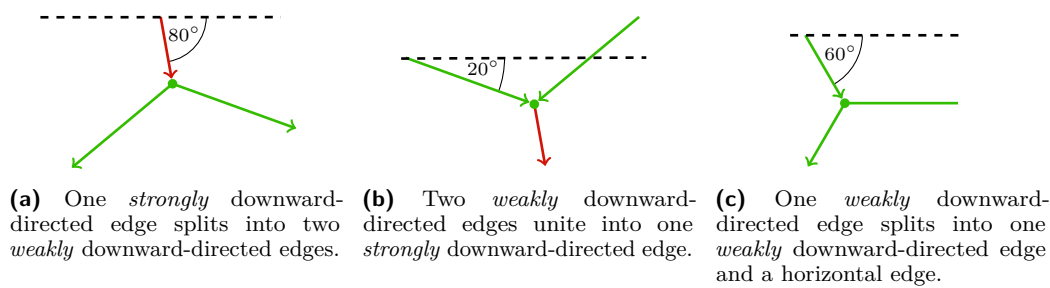
\begin{figure}[hbt]
    \centering
    \begin{subfigure}[t]{0.31\textwidth}
        \centering
        \begin{tikzpicture}[scale=1.8,rotate=10]

\coordinate (A) at (0,0);
\coordinate (B) at (0, 0.5);
\coordinate (C) at (0.866, -0.5);
\coordinate (D) at (-0.866, -0.5);

\coordinate (E) at (1, 0.5-0.174);
\pic[draw, angle eccentricity=1, angle radius=0.6cm] {angle=A--B--E};
\node[] at (0.15,0.35) {$\color{black}\scriptstyle\ang{80}$};

\draw [->,shorten >= 2pt,color={rgb, 255:red, 210; green, 20; blue, 7 }  ,draw opacity=1, line width=1] (B) -- (A);
\draw [->,shorten >= 2pt,color={rgb, 255:red, 52; green, 189; blue, 0 }  ,draw opacity=1, line width=1]    (A) -- (C);
\draw [->,shorten >= 2pt,color={rgb, 255:red, 52; green, 189; blue, 0 }  ,draw opacity=1, line width=1]    (A) -- (D);

\fill[fill={rgb, 255:red, 52; green, 189; blue, 0 }] (A) circle (1pt);

\draw[black, draw opacity=1, line width=1, dashed] (-0.866,0.65) -- (0.866,0.65-0.304) ;

\end{tikzpicture}
        \caption{One \emph{strongly} downward-directed edge splits into two \emph{weakly} downward-directed edges.}\label{fig:4lines:1}
    \end{subfigure}
    \hspace{2pt}
    \begin{subfigure}[t]{0.31\textwidth}
        \centering
        \begin{tikzpicture}[scale=1.8,rotate=10]

\coordinate (A) at (0,0);
\coordinate (B) at (0, -0.5);
\coordinate (C) at (0.866, 0.5);
\coordinate (D) at (-0.866, 0.5);

\coordinate (E) at (-0.866+1, 0.5-0.17);
\pic[draw, angle eccentricity=1, angle radius=1.3cm] {angle=A--D--E};
\node[] at (-0.32,0.33) {$\color{black}\scriptstyle\ang{20}$};

\draw [->,shorten >= 2pt,color={rgb, 255:red, 210; green, 20; blue, 7 }  ,draw opacity=1, line width=1] (A) -- (B);
\draw [->,shorten >= 2pt,color={rgb, 255:red, 52; green, 189; blue, 0 }  ,draw opacity=1, line width=1]    (C) -- (A);
\draw [->,shorten >= 2pt,color={rgb, 255:red, 52; green, 189; blue, 0 }  ,draw opacity=1, line width=1]    (D) -- (A);

\fill[fill={rgb, 255:red, 52; green, 189; blue, 0 }] (A) circle (1pt);

\draw[black, draw opacity=1, line width=1, dashed] (-0.9,0.5) -- (0.866,0.5-0.304);

\end{tikzpicture}
        \caption{Two \emph{weakly} downward-directed edges unite into one \emph{strongly} downward-directed edge.}\label{fig:4lines:2}
    \end{subfigure}
    \hspace{2pt}
    \begin{subfigure}[t]{0.31\textwidth}
        \centering
        \begin{tikzpicture}[scale=1.8]

\coordinate (A) at (0,0);
\coordinate (B) at (0.866, 0);
\coordinate (C) at (-0.289, 0.5);
\coordinate (D) at (-0.289, -0.5);

\coordinate (E) at (-0.289+1, 0.5);
\pic[draw, angle eccentricity=1, angle radius=0.75cm] {angle=A--C--E};
\node[] at (-0.03,0.37) {$\color{black}\scriptstyle\ang{60}$};

\draw [-,shorten >= 2pt,color={rgb, 255:red, 52; green, 189; blue, 0 },draw opacity=1, line width=1] (A) -- (B);
\draw [->,shorten >= 2pt,color={rgb, 255:red, 52; green, 189; blue, 0 },draw opacity=1, line width=1] (C) -- (A);
\draw [->,shorten >= 2pt,color={rgb, 255:red, 52; green, 189; blue, 0 } ,draw opacity=1, line width=1]    (A) -- (D);

\fill[fill={rgb, 255:red, 52; green, 189; blue, 0 }] (A) circle (1pt);

\draw[black, draw opacity=1, line width=1, dashed] (-0.4,0.5) -- (0.866,0.5);

\end{tikzpicture}
        \caption{One \emph{weakly} downward-directed edge splits into one \emph{weakly} downward-directed edge and a horizontal edge.}\label{fig:4lines:3}
    \end{subfigure}
    \caption{Illustration of the effect of Steiner points on the number of downward-directed edges. Red: \emph{strongly} downward-directed edges. Green: \emph{weakly} downward-directed edges.}
    \label{fig:4lines}
\end{figure}

    Observe that every edge connecting a point with $y>0$ to a point in $P$ is downward-directed. Since Steiner points always have an edge incoming from above, every downward-directed edge can be traced upward to a terminal. Therefore, to determine how many downward-directed edges can reach $P$, it suffices to consider the number of downward-directed edges originating from $q$ and $r$ and analyze how the addition of Steiner points affects this count.
    
    Because terminals also satisfy the angle condition, each terminal can give rise to either one strongly downward-directed edge or two weakly downward-directed edges. Steiner points, in turn, can influence the count of downward-directed edges reaching $P$ in the following ways:
    \begin{itemize}
        \item One strongly downward-directed edge splits into two weakly downward-directed edges, increasing the count by one (see Figure \ref{fig:4lines:1})
        \item Two weakly downward-directed edges unite into one strongly downward-directed edge, decreasing the count by one (see Figure \ref{fig:4lines:2}).
        \item One weakly downward-directed edge splits into one weakly downward-directed edge and a horizontal edge, leaving the count unchanged (see Figure \ref{fig:4lines:3}).
    \end{itemize}
    It follows that the maximum count of downward-directed edges reaching $P$ is achieved by splitting all strongly downward-directed edges into weakly downward-directed edges. Hence, with only two terminals above $y=0$, this count is at most $4$.
\end{proof}

Having shown that \textsc{FillHoles} terminates, we now demonstrate that it sufficiently reduces the number of holes. Since holes can only be introduced when the respective line points are connected through edges that do not lie on the line, we analyze paths running above the line to establish an upper bound of $10n$ on the number of holes, where $n$ denotes the number of real terminals.

\begin{lemma}\label{lemma:maxholes}
    After the termination of \textsc{FillHoles}, $T_\varepsilon$ contains at most $10n$ holes.
\end{lemma}
\begin{proof}
    Consider a connected component $C$ of the tree $T_\varepsilon$ excluding the line points, i.e., of $T_\varepsilon[V(T_\varepsilon)\setminus L]$. We will argue that if $C$ contains $k$ real terminals, then after termination of the \textsc{FillHoles} routine, there can be at most $10k$ holes between the leftmost and rightmost line points connected to this component in $T_\varepsilon$. We prove the lemma by summing over all these components.
    
    Let $l'_1$ be the leftmost and $l'_m$ the rightmost line points connected to $C$ in $T_\varepsilon$. Assume, without loss of generality, that $l'_1 \neq l'_m$; if there is only one connection, it is obvious that no holes appear. Between $l'_1$ and $l'_m$, there are
    $m-2$ line points and 
    $m-1$ line segments, spanning a distance of $(m-1)d$. We aim to show that there are at most $10k$ holes between $l'_1$ and $l'_m$. To analyze this, consider the path from $l'_1$ to $l'_m$ that, except for $l'_1$ and $l'_m$, only contains points from the component $C$. 
    After Step~3 of \textsc{FillHoles}, all edges on this path must have length less than $d$, as we will first prove:
    Suppose there exists an edge $e$ on the path with $c(e) \geq d$. As $l'_1$ and $l'_m$ are the only line points on the path, $e$ is not a line segment. Removing $e$ from $T_{\epsilon}$ would split $T_\varepsilon$ into two connected components, with $l'_1$ and $l'_m$ residing in different connected components. Since every line point in between belongs to either one or the other connected component, there must exist two consecutive line points, $l'_i$ and $l'_{i+1}$, that lie in different components. The edge $\{l'_i, l'_{i+1}\}$ would reconnect these two components and has length $d$. Thus, $e$ would have been replaced by $\{l'_i, l'_{i+1}\}$ in Step~3, contradicting our assumption. Therefore, all edges on the path from $l'_1$ to $l'_m$ are shorter than $d$. This implies that this path must contain at least $m$ edges.

    Now, consider the subpath $P = (p_1, \dots, p_b)$ with $\{l'_1,p_1\}\in E(T_\varepsilon)$ and $\{p_b,l'_m\}\in  E(T_\varepsilon)$ and $p_i\in V(C)$ for $i\in \{1,\dots,b\}$. Observe that $P$ contains at least $m-2$ edges, so we can write $b \geq m-1$. We partition $P$ into consecutive blocks of size 5, ignoring any leftover points. This results in exactly $\left\lfloor\frac{b}{5} \right\rfloor$ blocks. After the termination of \textsc{FillHoles}, each block must contain at least one point that is either a real terminal or does not share an edge with a line point. Otherwise, Step~4 would have identified five consecutive Steiner points, all sharing an edge with a line point, and the algorithm would have replaced them with a local optimum, destroying this specific substructure. Therefore, we know that each block contains either a real terminal or it contains a Steiner point $s$ that does not share an edge with a line point. In the latter case, we examine the subtree we examine the subtree branching off from $P$ at $s$. This subtree always exists and is unique because, after Step~2, all Steiner points have degree 3. Since $s$ does not share an edge with a line point, this subtree must either contain more than one line point or a real terminal.
    As $T_\varepsilon$ is a tree, the subtrees branching off from $P$ are disjoint.
    
    Let
    \begin{itemize}
        \item $\rho$ be the number of points in $P$ that are either real terminals or are Steiner points that have a real terminal in their branching-off subtree,  
        \item $\tau$ be the number of Steiner points in $P$ that share an edge with a line point,  
        \item $\pi$ be the number of Steiner points in $P$ that have no real terminal in their subtree, but at least two line points.
    \end{itemize}
     Since every point on $P$ belongs to exactly one of the three classes, it holds that $\pi + \tau + \rho = b$. An example of this classification is shown in Figure \ref{fig:maxholes:example}.
    \begin{figure}[htb]
    \centering
\definecolor{darkgreen}{rgb}{0.0, 0.7, 0.0}

\pgfdeclarelayer{nodelayer}
\pgfdeclarelayer{edgelayer}
\pgfsetlayers{edgelayer,nodelayer,main}
\tikzstyle{terminal}=[fill=black, draw=black, shape=circle, inner sep=1.5pt]
\tikzstyle{tau}=[fill={rgb, 255:red, 52; green, 189; blue, 0 }, draw={rgb, 255:red, 52; green, 189; blue, 0 }, shape=circle, inner sep=1.5pt]
\tikzstyle{pi}=[fill=blue, draw=blue, shape=circle, inner sep = 1.5pt]
\tikzstyle{rho}=[fill={rgb,255: red,236; green,0; blue,0}, draw={rgb,255: red,227; green,0; blue,0}, shape=circle, inner sep=1.5pt]
\tikzstyle{line}=[fill=white, draw=black, shape=circle, inner sep=1.5pt,line width=1]
\tikzstyle{steiner}=[fill={rgb,255:red,160; green,160; blue,160}, draw={rgb,255:red,160; green,160; blue,160},shape=circle,line width=1, inner sep=1.5pt]

\tikzstyle{new edge style 0}=[-, fill=none, line width=1]
\tikzstyle{pathedge}=[-, fill=white, line width=1, draw={rgb, 255:red, 52; green, 189; blue, 0 }]
\begin{tikzpicture}
	\begin{pgfonlayer}{nodelayer}
		\node [style=line] (30) at (0, -3) {};
                \node [left] at (30.west) {$l'_1$};

		\node [style=line] (31) at (1, -3) {};
		\node [style=line] (32) at (2, -3) {};
		\node [style=line] (33) at (3, -3) {};
		\node [style=line] (34) at (4, -3) {};
		\node [style=line] (35) at (5, -3) {};
		\node [style=line] (36) at (6, -3) {};
		\node [style=line] (37) at (7, -3) {};
		\node [style=line] (38) at (8, -3) {};
		\node [style=line] (39) at (9, -3) {};
                \node [right] at (39.east) {$l'_{10}$};
  
		\node [style=tau] (40) at (0.75, -2.5) {};
		\node [style=tau] (41) at (5.25, -2.25) {};
		\node [style=rho] (46) at (1.5, -2) {};
		\node [style=pi] (47) at (3.5, -1.75) {};
		\node [style=rho] (48) at (4.25, -2) {};
		\node [style=terminal] (49) at (4.75, -0.75) {};
		\node [style=rho] (50) at (6, -1.75) {};
		\node [style=steiner] (51) at (3.25, -2.25) {};
		\node [style=steiner] (52) at (2.5, -2.5) {};
		\node [style=tau] (53) at (8.5, -2.5) {};
		\node [style=steiner] (54) at (6.5, 1.25) {};
		\node [style=terminal] (55) at (6, 2.5) {};
		\node [style=terminal] (56) at (7.25, 2.25) {};
		\node [style=rho] (59) at (7, -1.75) {};
		\node [style=terminal] (60) at (6.25, -2.25) {};
		\node [style=steiner] (61) at (6.75, -2.25) {};
		\node [style=rho] (62) at (2.5, -1.75) {};
		\node [style=terminal] (63) at (2.5, -0.5) {};
		\node [style=rho] (64) at (7.75, -2) {};
		\node [style=terminal] (65) at (8.75, -0.75) {};
	\end{pgfonlayer}
	\begin{pgfonlayer}{edgelayer}
		\draw [style=pathedge] (50) to (41);
		\draw [style=pathedge] (41) to (48);
		\draw [style=pathedge] (48) to (47);
		\draw [style=pathedge] (46) to (40);
		\draw [style=new edge style 0] (40) to (30);
		\draw [style=new edge style 0] (40) to (31);
		\draw [style=new edge style 0] (32) to (52);
		\draw [style=new edge style 0] (52) to (33);
		\draw [style=new edge style 0] (52) to (51);
		\draw [style=new edge style 0] (51) to (34);
		\draw [style=new edge style 0] (36) to (37);
		\draw [style=new edge style 0] (51) to (47);
		\draw [style=new edge style 0] (49) to (48);
		\draw [style=new edge style 0] (41) to (35);
		\draw [style=new edge style 0] (39) to (53);
		\draw [style=new edge style 0] (38) to (53);
		\draw [style=new edge style 0] (55) to (54);
		\draw [style=new edge style 0] (54) to (56);
		\draw [style=new edge style 0] (54) to (50);
		\draw [style=new edge style 0] (60) to (61);
		\draw [style=new edge style 0] (61) to (37);
		\draw [style=pathedge] (50) to (59);
		\draw [style=new edge style 0] (61) to (59);
		\draw [style=pathedge] (46) to (62);
		\draw [style=pathedge] (62) to (47);
		\draw [style=new edge style 0] (63) to (62);
		\draw [style=pathedge] (64) to (53);
		\draw [style=pathedge] (64) to (59);
		\draw [style=new edge style 0] (64) to (65);
	\end{pgfonlayer}
\end{tikzpicture}
    \caption{Example illustrating Lemma \ref{lemma:maxholes}. The path $P$ is represented by green edges. Green points are Steiner points on $P$ that share an edge with a line point \hbox{($\tau = 3$)}. Red points are real terminals on $P$, or are Steiner points on $P$ with at least one real terminal in their branching-off subtree ($\rho = 6$). The blue point represents a Steiner point on the path with at least two line points and no real terminal in its subtree ($\pi= 1$). Remaining Steiner points, terminals and line points not on the path $P$ are gray, black and white, respectively. In this example, $m=b=10$.}
    \label{fig:maxholes:example}
\end{figure}
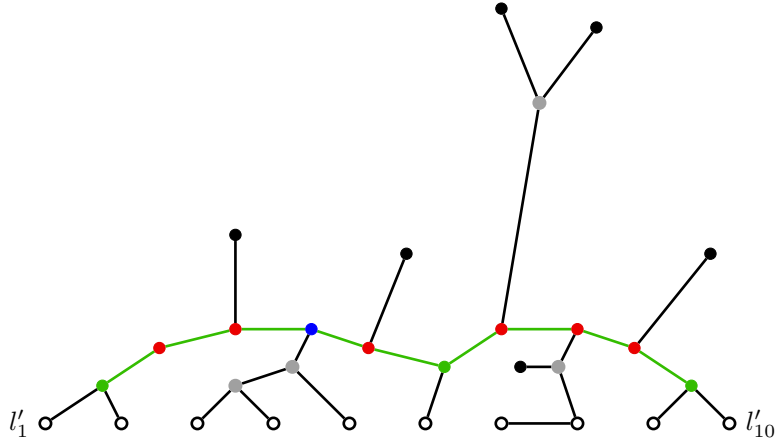

From these definitions, we make the following observations: 
   
\begin{equation}
     \rho + \pi \geq \left\lfloor\frac{b}{5} \right\rfloor \label{eq:eachblockatleastone}
\end{equation}
as each of the $\left\lfloor\frac{b}{5} \right\rfloor$ blocks contains at least one of the following: a real terminal, a Steiner point with a real terminal in its subtree, or a Steiner point with two line points in its subtree.
\begin{equation}
    k \geq \rho \label{eq:fewterminals}
\end{equation}
as for each Steiner point on $P$ with a terminal in its branching-off subtree, there must exist at least one real terminal (there are $k$ real terminals in the component). These subtree-terminals are distinct as the subtrees branching off from $P$ are disjoint.

\begin{equation}
     m-2 \geq 2\pi + \tau \label{eq:fewlinepoints}
\end{equation}
since there are only $m-2$ line points between $l'_1$ and $l'_m$, and the subtrees branching-off from $P$ are disjoint. Using $b \geq m-1$, we deduce
    \[
        \pi + \tau + \rho = b \geq m-1 \stackrel{\mathclap{(\ref{eq:fewlinepoints})}}{\geq} 2\pi + \tau + 1 \quad \implies \quad \rho \geq \pi + 1.
    \]
    Using this, we derive  
    \[
        2\rho \geq \rho + \pi + 1 \stackrel{\mathclap{(\ref{eq:eachblockatleastone})}}{\geq} \left\lfloor \frac{b}{5} \right\rfloor + 1 \quad \implies \quad \rho \geq \frac{1}{2}\left(\left\lfloor \frac{b}{5} \right\rfloor + 1\right),
    \]  
    which simplifies to 
    \[
        \rho \geq \frac{1}{2} \left\lfloor \frac{b+5}{5} \right\rfloor \geq \frac{1}{2} \left\lfloor \frac{m+4}{5} \right\rfloor \geq \frac{1}{2} \frac{m}{5} = \frac{m}{10}.
    \]
    Since there can be at most $m-1$ holes between $l'_1$ and $l'_m$, it follows that when a component has $k$ real terminals, there can be at most $m-1 < m \leq 10\rho \stackrel{\mathclap{(\ref{eq:fewterminals})}}{\leq} 10k$ holes between their leftmost and rightmost connected line point.
    
    Note that holes can only be introduced when the respective line points are connected through edges that do not lie on the line, i.e., are part of a component. Because components are disjoint, they do not share any real terminals. Thus, we can sum over all components to obtain an upper bound of $10n$ on the number of holes in $T_\varepsilon$, where $n$ is the number of real terminals in~$T_\varepsilon$.
\end{proof}

Now that we have shown that \textsc{FillHoles} sufficiently reduces the number of holes, we can combine all the results and prove that Algorithm~\ref{alg:ptas} retains approximation guarantees for the \esfl. Specifically, we will argue that introducing the zero-cost line to the \est-instance reduces the solution cost of $T_\varepsilon$ by almost the full width of the instance $W_I$. Since $OPT_{I_\varepsilon} - W_I$ provides a lower bound on the \esfl\ solution value, this yields a tight approximation factor.

\ptas*

\begin{proof}
    Given an instance $I$ and a parameter $\varepsilon > 0$, we will first show that Algorithm \ref{alg:ptas} runs in polynomial time for a fixed $\varepsilon$. 
    Using the PTAS of Kisfaludi-Bak et al.~\cite{kisfaludi2021:ptas} we can achieve Step~1, namely a $(1+\varepsilon)$-approximation to the \est-instance $I_\varepsilon$ with $n+\ceil{\frac{n}{\varepsilon}}+1=O(\frac{n}{\varepsilon})$ terminals, in 
    $O\left(2^{O(\frac{1}{\varepsilon})}\cdot \frac{n}{\varepsilon} + \text{poly}(\frac{1}{\varepsilon}) \frac{n}{\varepsilon} \log(\frac{n}{\varepsilon})\right)$.
    The \textsc{FillHoles} Subroutine (Algorithm \ref{alg:transform}) has a runtime of $O((\frac{n}{\varepsilon})^3)$, as shown in Lemma \ref{lemma:runtime}. Since Step 4 and 5 of Algorithm \ref{alg:ptas} can easily be implemented in time $O((\frac{n}{\varepsilon})^2)$, we achieve a total runtime of $O\left(2^{O(\frac{1}{\varepsilon})}\cdot \frac{n}{\varepsilon} + \text{poly}(\frac{1}{\varepsilon}) \frac{n}{\varepsilon} \log(\frac{n}{\varepsilon}) + (\frac{n}{\varepsilon})^3\right)$.
    
    It remains to analyze the approximation guarantee. Denote the cost of the solution $T_\varepsilon$ after the termination of \textsc{FillHoles} by $c(T_\varepsilon)$, the cost of the final solution $T$ by $c(T)$, and the costs of the respective optimal solutions by $OPT_{I_\varepsilon}$ and $OPT_I$.
    It holds that
    \begin{equation} \label{eq}
        OPT_{I_\varepsilon} \leq OPT_I + W_I
    \end{equation}
    because an optimal solution $T^*$ for $I$ can be transformed into a valid solution for $I_\varepsilon$ by adding all edges along the line and removing any cycles. This is true since $T^*$ is a tree containing all real terminals and at least one point on the line. Thus, the cost of $T^*$ increases at most by the cost of all line segments, which is $W_I$.
   
    For the distance between line points $d$, it holds that
    \[
    d=\frac{W_I}{\ceil{\frac{n}{\varepsilon}}} \leq \frac{W_I}{\frac{n}{\varepsilon}} = \frac{\varepsilon}{n}\cdot W_I.
    \]
    
    Let $W' = c(T_\varepsilon) - c(T)$ denote the cost savings resulting from adding the cost-free line $\gamma$ to the tree $T_\varepsilon$, replacing all line segments. According to Lemma~\ref{lemma:maxholes}, there are at most $10n$ holes in $T_\varepsilon$, which implies
    \[
    W' \geq W_I - 10nd  \geq W_I - 10n\frac{\varepsilon}{n}W_I = (1- 10\varepsilon)\cdot W_I.
    \]
    Therefore, it holds that
    \begin{align*}
    c(T)& =\quad c(T_\varepsilon) - W' \\[1mm]
    &\leq \quad (1+\varepsilon) OPT_{I_\varepsilon}-(1-10\varepsilon) W_I \\[1mm]
    & = \quad OPT_{I_\varepsilon} + \varepsilon OPT_{I_\varepsilon} - W_I + 10 \varepsilon  W_I \\[1mm]
    & \stackrel{\mathclap{(\ref{eq})}}{\leq} \quad OPT_{I} + \varepsilon  OPT_{I_\varepsilon} + 10 \varepsilon  W_I \\[1mm]
    & \stackrel{\mathclap{(\ref{eq})}}{\leq} \quad OPT_{I} + \varepsilon (W_I + OPT_I) + 10 \varepsilon   W_I \\[1mm]
    & = \quad OPT_{I} + \varepsilon OPT_I + 11 \varepsilon   W_I \\[1mm]
    & \stackrel{\mathclap{Lemma~\ref{lowerBound}}}{\leq}\quad OPT_I + \varepsilon OPT_I + \left(11+\frac{22}{\tan(\ang{30})}\right) \varepsilon OPT_I \\[1mm]
     & \leq \quad(1+50.106  \varepsilon) \cdot OPT_I\\[1mm]
     & \leq \quad(1+51  \varepsilon) \cdot OPT_I.\\[1mm]
\end{align*}
Although this approach yields a $(1+51 \varepsilon)$-approximation instead of a $(1+\varepsilon)$-approximation, we can easily adjust the algorithm as follows: Given the parameter $\varepsilon$, we define $\varepsilon' := \varepsilon / 51$ and use Algorithm \ref{alg:ptas} to achieve a $(1+51 \varepsilon')$=$(1+\varepsilon)$-approximation in
\begin{align*}
&O\left(2^{O(\frac{1}{\varepsilon'})}\cdot \frac{n}{\varepsilon'} + \text{poly}\left(\frac{1}{\varepsilon'}\right) \frac{n}{\varepsilon'} \log\left(\frac{n}{\varepsilon'}\right) + \left(\frac{n}{\varepsilon'}\right)^3\right)\\
= \text{ }& O\left(2^{O(\frac{51}{\varepsilon})}\cdot \frac{51n}{\varepsilon} + \text{poly}\left(\frac{51}{\varepsilon}\right) \frac{51n}{\varepsilon} \log\left(\frac{51n}{\varepsilon}\right) + \left(\frac{51n}{\varepsilon}\right)^3\right)\\
= \text{ }& O\left(2^{O(\frac{1}{\varepsilon})}\cdot \frac{n}{\varepsilon} + \text{poly}\left(\frac{1}{\varepsilon}\right) \frac{n}{\varepsilon} \log\left(\frac{n}{\varepsilon}\right) + \left(\frac{n}{\varepsilon}\right)^3\right),
\end{align*}
 where the factor of $51$ disappears in the constants of the $O$-notation.
\end{proof}
Together with Lemma \ref{lemma:li} and Theorem \ref{thm:ptas}, we can also settle the existence of a PTAS for the \esl.
\ptascor*
\begin{proof}
    According to Lemma \ref{lemma:li}, there always exists an optimum solution to the \esl\ in which $\gamma$ passes through at least two terminal points. Therefore, it suffices to apply our \esfl-PTAS to each of the $\binom{n}{2}$ \esfl-instances where the line passes through two distinct terminal points and select the best resulting solution. This results in an additional runtime factor of $n^2$.
\end{proof}

\section{Conclusion}
In this work, we established the NP-hardness of both the Euclidean Steiner Line Problem (\esl) and the Euclidean Steiner fixed Line Problem (\esfl). Additionally, we proved that both problems admit a polynomial-time approximation scheme (PTAS), by employing a PTAS-reduction to the classical Euclidean Steiner Tree Problem (\est). Using the fastest currently known PTAS for the \est\ by Kisfaludi-Bak et al.~\cite{kisfaludi2021:ptas}, we obtain a runtime of $O\left(2^{O(\frac{1}{\varepsilon})}\cdot \frac{n}{\varepsilon} + \textnormal{poly}\left(\frac{1}{\varepsilon}\right) \frac{n}{\varepsilon} \log\left(\frac{n}{\varepsilon}\right) + \left(\frac{n}{\varepsilon}\right)^3\right)$ to compute a $(1+\varepsilon)$-approximation to the \esfl.

Further research could include improving the runtime of the PTAS. Specifically, we believe that by developing a more sophisticated \textsc{FillHoles} routine, one could reduce its cubic dependence on the number of terminals.

Another promising direction for future work is generalizing our results to the problem variant introduced by Althaus et al.~\cite{althaus2020}, where terminals can be line segments of arbitrary length. A key challenge in this context would be to establish a sufficient lower bound on the optimal solution, analogous to Lemma \ref{lowerBound}. Furthermore, it would be interesting to analyze the  complexity of the Steiner Line Problem, defined on other uniform orientation metrics such as the rectilinear or the octilinear metric. One could also investigate adapting the PTAS to these metrics.



\bibliography{steiner.bib}

\begin{thebibliography}{10}

\bibitem{althaus2020}
Ernst Althaus, Felix Rauterberg, and Sarah Ziegler.
\newblock {Computing Euclidean Steiner trees over segments}.
\newblock {\em {EURO} J. Comput. Optim.}, 8(3):309--325, 2020.
\newblock URL: \url{https://doi.org/10.1007/s13675-020-00125-w}, \href
  {https://doi.org/10.1007/S13675-020-00125-W}
  {\path{doi:10.1007/S13675-020-00125-W}}.

\bibitem{arora:ptas}
Sanjeev Arora.
\newblock {Polynomial Time Approximation Schemes for Euclidean Traveling
  Salesman and other Geometric Problems}.
\newblock {\em J. {ACM}}, 45(5):753--782, 1998.
\newblock \href {https://doi.org/10.1145/290179.290180}
  {\path{doi:10.1145/290179.290180}}.

\bibitem{bose2020}
Prosenjit Bose, Anthony D'Angelo, and Stephane Durocher.
\newblock {On the Restricted 1-Steiner Tree Problem}.
\newblock In Donghyun Kim, R.~N. Uma, Zhipeng Cai, and Dong~Hoon Lee, editors,
  {\em Computing and Combinatorics - 26th International Conference, {COCOON}
  2020, Atlanta, GA, USA, August 29-31, 2020, Proceedings}, volume 12273 of
  {\em Lecture Notes in Computer Science}, pages 448--459. Springer, 2020.
\newblock \href {https://doi.org/10.1007/978-3-030-58150-3\_36}
  {\path{doi:10.1007/978-3-030-58150-3\_36}}.

\bibitem{chung1985}
F.~R.~K. Chung and R.~L. Graham.
\newblock {A new bound for Euclidean Steiner minimal trees}.
\newblock {\em Annals of the New York Academy of Sciences}, 440(1):328--346,
  1985.
\newblock URL:
  \url{https://nyaspubs.onlinelibrary.wiley.com/doi/abs/10.1111/j.1749-6632.1985.tb14564.x},
  \href {https://doi.org/10.1111/j.1749-6632.1985.tb14564.x}
  {\path{doi:10.1111/j.1749-6632.1985.tb14564.x}}.

\bibitem{cui2021}
Ziyuan Cui, Hai Lin, Yan Wu, Yufei Wang, and Xiao Feng.
\newblock {Optimization of Pipeline Network Layout for Multiple Heat Sources
  Distributed Energy Systems Considering Reliability Evaluation}.
\newblock {\em Processes}, 9(8), 2021.
\newblock URL: \url{https://www.mdpi.com/2227-9717/9/8/1308}, \href
  {https://doi.org/10.3390/pr9081308} {\path{doi:10.3390/pr9081308}}.

\bibitem{garey1977}
M.~R. Garey, R.~L. Graham, and D.~S. Johnson.
\newblock {The Complexity of Computing Steiner Minimal Trees}.
\newblock {\em SIAM J. Appl. Math.}, 32(4):835–859, June 1977.
\newblock \href {https://doi.org/10.1137/0132072} {\path{doi:10.1137/0132072}}.

\bibitem{gilbertpollak1968}
E.~N. Gilbert and H.~O. Pollak.
\newblock {Steiner Minimal Trees}.
\newblock {\em SIAM Journal on Applied Mathematics}, 16(1):1--29, 1968.
\newblock \href {https://doi.org/10.1137/0116001} {\path{doi:10.1137/0116001}}.

\bibitem{holby2017}
Jack Holby.
\newblock {{Variations on the Euclidean Steiner Tree Problem and Algorithms}}.
\newblock {\em Rose-Hulman Undergraduate Mathematics Journal}, 18(1):Article 7,
  2017.

\bibitem{kisfaludi2021:ptas}
S{\'{a}}ndor Kisfaludi{-}Bak, Jesper Nederlof, and Karol Wegrzycki.
\newblock {A Gap-ETH-Tight Approximation Scheme for Euclidean {TSP}}.
\newblock In {\em 62nd {IEEE} Annual Symposium on Foundations of Computer
  Science, {FOCS} 2021, Denver, CO, USA, February 7-10, 2022}, pages 351--362.
  {IEEE}, 2021.
\newblock \href {https://doi.org/10.1109/FOCS52979.2021.00043}
  {\path{doi:10.1109/FOCS52979.2021.00043}}.

\bibitem{li2024:segments}
Jian-Ping Li, Su-Ding Liu, Jun-Ran Lichen, Peng-Xiang Pan, and Wen-Cheng Wang.
\newblock {Approximation Algorithms for Solving the 1-Line Minimum Steiner Tree
  of Line Segments Problem}.
\newblock {\em Journal of the Operations Research Society of China}, 12(3):729,
  2024.
\newblock URL:
  \url{https://www.jorsc.shu.edu.cn/EN/abstract/article_20657.shtml}, \href
  {https://doi.org/10.1007/s40305-022-00437-1}
  {\path{doi:10.1007/s40305-022-00437-1}}.

\bibitem{li2022:rectilinear}
Jianping Li, Junran Lichen, Wencheng Wang, Jean Yeh, Yeong{-}Nan Yeh, Xingxing
  Yu, and Yujie Zheng.
\newblock {1-line minimum rectilinear steiner trees and related problems}.
\newblock {\em J. Comb. Optim.}, 44(4):2832--2852, 2022.
\newblock URL: \url{https://doi.org/10.1007/s10878-021-00796-0}, \href
  {https://doi.org/10.1007/S10878-021-00796-0}
  {\path{doi:10.1007/S10878-021-00796-0}}.

\bibitem{li2023:bottleneck}
Jianping Li, Suding Liu, and Junran Lichen.
\newblock {An Exact Algorithm for the Line-Constrained Bottleneck k-Steiner
  Tree Problem}.
\newblock In Weili Wu and Jianxiong Guo, editors, {\em Combinatorial
  Optimization and Applications - 17th International Conference, {COCOA} 2023,
  Hawaii, HI, USA, December 15-17, 2023, Proceedings, Part {I}}, volume 14461
  of {\em Lecture Notes in Computer Science}, pages 434--445. Springer, 2023.
\newblock \href {https://doi.org/10.1007/978-3-031-49611-0\_31}
  {\path{doi:10.1007/978-3-031-49611-0\_31}}.

\bibitem{li2020:line}
Jianping Li, Suding Liu, Junran Lichen, Wencheng Wang, and Yujie Zheng.
\newblock {Approximation algorithms for solving the 1-line Euclidean minimum
  Steiner tree problem}.
\newblock {\em J. Comb. Optim.}, 39(2):492--508, 2020.
\newblock URL: \url{https://doi.org/10.1007/s10878-019-00492-0}, \href
  {https://doi.org/10.1007/S10878-019-00492-0}
  {\path{doi:10.1007/S10878-019-00492-0}}.

\bibitem{li2021:minsteiner}
Jianping Li, Yujie Zheng, Junran Lichen, and Wencheng Wang.
\newblock {On the minimum number of Steiner points of constrained 1-line-fixed
  Steiner tree in the Euclidean plane $\mathbb{R}^2$}.
\newblock {\em Optim. Lett.}, 15(2):669--683, 2021.
\newblock URL: \url{https://doi.org/10.1007/s11590-020-01627-7}, \href
  {https://doi.org/10.1007/S11590-020-01627-7}
  {\path{doi:10.1007/S11590-020-01627-7}}.

\bibitem{liu2024:minsteiner}
Suding Liu.
\newblock {Approximation algorithm for solving the 1-line Steiner tree problem
  with minimum number of Steiner points}.
\newblock {\em Optim. Lett.}, 18(6):1421--1435, 2024.
\newblock URL: \url{https://doi.org/10.1007/s11590-023-02058-w}, \href
  {https://doi.org/10.1007/S11590-023-02058-W}
  {\path{doi:10.1007/S11590-023-02058-W}}.

\bibitem{melzak1961}
Zdzislaw~Alexander Melzak.
\newblock {On the problem of Steiner}.
\newblock {\em Canadian Mathematical Bulletin}, 4(2):143--148, 1961.

\bibitem{mitchell1999:ptas}
Joseph S.~B. Mitchell.
\newblock {Guillotine Subdivisions Approximate Polygonal Subdivisions: {A}
  Simple Polynomial-Time Approximation Scheme for Geometric TSP, k-MST, and
  Related Problems}.
\newblock {\em {SIAM} J. Comput.}, 28(4):1298--1309, 1999.
\newblock \href {https://doi.org/10.1137/S0097539796309764}
  {\path{doi:10.1137/S0097539796309764}}.

\bibitem{pollak1978}
Henry~O. Pollak.
\newblock {Some Remarks on the Steiner Problem}.
\newblock {\em J. Comb. Theory {A}}, 24(3):278--295, 1978.
\newblock \href {https://doi.org/10.1016/0097-3165(78)90058-4}
  {\path{doi:10.1016/0097-3165(78)90058-4}}.

\bibitem{raosmith:ptas}
Satish Rao and Warren~D. Smith.
\newblock {Approximating Geometrical Graphs via "Spanners" and "Banyans"}.
\newblock In Jeffrey~Scott Vitter, editor, {\em Proceedings of the Thirtieth
  Annual {ACM} Symposium on the Theory of Computing, Dallas, Texas, USA, May
  23-26, 1998}, pages 540--550. {ACM}, 1998.
\newblock \href {https://doi.org/10.1145/276698.276868}
  {\path{doi:10.1145/276698.276868}}.

\bibitem{rubinstein1997}
J.~Hyam Rubinstein, Doreen~A. Thomas, and Nicholas~C. Wormald.
\newblock {Steiner Trees for Terminals Constrained to Curves}.
\newblock {\em {SIAM} J. Discret. Math.}, 10(1):1--17, 1997.
\newblock \href {https://doi.org/10.1137/S0895480192241190}
  {\path{doi:10.1137/S0895480192241190}}.

\bibitem{geosteiner96}
Pawel Winter and Martin Zachariasen.
\newblock {Euclidean Steiner minimum trees: An improved exact algorithm}.
\newblock {\em Networks}, 30(3):149--166, 1997.
\newblock \href
  {https://doi.org/10.1002/(SICI)1097-0037(199710)30:3<149::AID-NET1>3.0.CO;2-L}
  {\path{doi:10.1002/(SICI)1097-0037(199710)30:3<149::AID-NET1>3.0.CO;2-L}}.

\end{thebibliography}

\newpage

\appendix

\section*{Appendix}\label{appendix:observation}

\observationCases*
\begin{proof}

    Consider $g(a):=|5a+3|+|3a+5|-\frac{5}{2}\sqrt{a^2+1}$.
    We use a case distinction to show $g(a)>0$. First, note that $g(-\frac{5}{3})>0$ and $g(-\frac{3}{5})>0$.

    \begin{itemize}
        \item 
    For $a<-\frac{5}{3}$, $g$ is given by $g(a)=-5a-3-3a-5-\frac{5}{2}\sqrt{a^2+1}$. For the first derivative of $g$, it holds that \[\frac{dg}{da} = -8-\frac{5a}{2\sqrt{a^2+1}} \leq -8 + \left|\frac{5a}{2\sqrt{a^2+1}}\right| \leq -8+\frac{5}{2}<0.\] Since $g(-\frac{5}{3})>0$, it follows that $g(a)>0$ for $a < -\frac{5}{3}$.

    \item 
    For $a>-\frac{3}{5}$, $g$ is given by $g(a)=5a+3+3a+5-\frac{5}{2}\sqrt{a^2+1}$. For the first derivative of $g$, it holds that \[\frac{dg}{da} = 8-\frac{5a}{2\sqrt{a^2+1}} \geq 8 - \left|\frac{5a}{2\sqrt{a^2+1}}\right| \geq 8-\frac{5}{2}>0.\] Since $g(-\frac{3}{5})>0$, it follows that $g(a)>0$ for $a < -\frac{3}{5}$.

    \item 
    Finally, for $-\frac{5}{3}\leq a\leq -\frac{3}{5}$, $g$ is given by $g(a)=-5a-3+3a+5-\frac{5}{2}\sqrt{a^2+1}$. For the second derivative of $g$, it holds that \[\frac{d^2g}{da^2} = -\frac{5}{2(a^2+1)^\frac{3}{2}}<0.\] Since $g$ is concave and both $g(-\frac{5}{3})>0$ and $g(-\frac{3}{5})>0$, it follows that $g(a)>0$ for $-\frac{5}{3}\leq a\leq -\frac{3}{5}$.

    \end{itemize}

\end{proof}

\end{document}